\documentclass[structabstract,fleqn]{aa}  
\usepackage{cases}
\usepackage{graphicx}
\usepackage{natbib}
\bibpunct{(}{)}{;}{a}{}{,} % to follow the A&A style
\usepackage{txfonts}
\begin{document}
   \title{Milky Way mass models for orbit calculations}

%   \subtitle{}

   \author{A.~Irrgang\inst{\ref{remeis}}
           \and
           B.~Wilcox\inst{\ref{remeis},\ref{standrews}}
           \and
           E.~Tucker\inst{\ref{remeis},\ref{tucson}}
           \and
           L.~Schiefelbein\inst{\ref{remeis},\ref{MIT}}
           }

\institute{Dr.~Karl~Remeis-Observatory \& ECAP, Astronomical Institute, Friedrich-Alexander University Erlangen-Nuremberg,\\ Sternwartstr.~7, 96049 Bamberg, Germany\\
           \email{andreas.irrgang@sternwarte.uni-erlangen.de}\label{remeis}
           \and
           School of Physics and Astronomy, University of St.~Andrews, North Haugh, St.~Andrews, KY16 9SS, UK\label{standrews}
           \and
           Department of Physics, University of Arizona, Tucson, AZ 85721, USA\label{tucson}
           \and
           Massachusetts Institute of Technology, Cambridge, MA 02139, USA\label{MIT}           
          }

   \date{Received 11 October 2012 / Accepted 16 November 2012}

% \abstract{}{}{}{}{} 
% 5 {} token are mandatory
 
  \abstract
   {Studying the trajectories of objects like stars, globular clusters, or satellite galaxies in the Milky Way allows the dark matter halo to be traced but requires reliable models of its gravitational potential.}
   {Realistic, yet simple and fully analytical, models have already been presented in the past. However, improved, as well as new, observational constraints have become available in the meantime, calling for a recalibration of the respective model parameters.}
   {Three widely used model potentials are revisited. By a simultaneous least-squares fit to the observed rotation curve, in-plane proper motion of Sgr~A*, local mass/surface density, and the velocity dispersion in Baade's window, parameters of the potentials are brought up-to-date. The mass at large radii --~in particular, that of the dark matter halo~-- is hereby constrained by requiring that the most extreme known halo, blue horizontal-branch star has to be bound to the Milky Way.}
   {The Galactic mass models are tuned to yield a very good match to recent observations. The mass of the dark matter halo is -- within the limitations of the applied models -- estimated in a fully consistent way. As a first application, the trajectory of the hypervelocity star \object{HE~0437-5439} is investigated again to check its suggested origin in the Large Magellanic Cloud (LMC).}
   {Despite their simplicity, the presented Milky Way mass models are very able to reproduce all observational constraints. Their analytical and simple form makes them ideally suited for fast and accurate orbit calculations. The LMC cannot be ruled out as HE~0437-5439's birthplace.}


   \maketitle

\section{Introduction}
Tracking large-scale orbits within the Milky Way is an important component in the study of the dynamical properties of stars, globular clusters, or satellite galaxies. Given the object's coordinates and velocity components, this is a straightforward task once the Galactic gravitational potential is available. Owing to a lack of observational constraints, a great deal of the Milky Way's mass distribution --~in particular of its dark matter halo~-- remains unknown. Therefore, different mass models are able to reproduce the observations equally well. For the purpose of numerical orbit calculations, a mathematically simple and analytically closed potential is preferred because it supports fast computations, which is especially favorable when using Monte Carlo methods for error estimation.

The Galactic gravitational potential of \citet{Allen_Santillan} perfectly fulfills these criteria and has thus been widely used to calculate trajectories for, e.g., globular clusters \citep{odenkirchen_brosche, allen_moreno_pichardo, lane_etal}, dwarf spheroidals \citep{lepine_etal}, planetary nebulae \citep{wu_etal}, white dwarfs \citep{pauli1,pauli2}, horizontal-branch stars \citep{altmann_deboer,kaempf_deboer_altmann}, subluminous B stars \citep{altmann_edelmann_deboer,tillich_etal}, halo stars \citep{schuster_et_al,pereira_etal}, runaway stars \citep{irrgang_etal,silva_napiwotzki}, and hypervelocity stars \citep{hirsch_etal,Edelmann_etal}. However, the model parameters can be recalibrated using new and improved observational constraints. In this context, a simplified expression for the halo component is introduced as well.

As for similar studies, standard constraints encompassing terminal velocities for the inner rotation curve, circular velocities of maser sources for the outer rotation curve, in-plane proper motion of Sgr~A*, local mass/surface density, and the velocity dispersion in Baade's window are fitted to determine the respective parameters. To constrain the mass at large radii, in particular that of the dark matter halo, the approach of \citet{J15390239} is followed here by requiring the dynamically peculiar halo star \object{SDSSJ153935.67+023909.8} (J1539+0239 for short) to be gravitationally bound to the Milky Way (Sect.~\ref{ObservationsandFitting}).

Despite its simplicity, the resulting revised \citeauthor{Allen_Santillan} model (hereafter denoted Model~I) is capable of meeting all of its imposed conditions. In addition to the fast computation of realistic trajectories, calibration of the potential using the star J1539+0239 allows the halo mass to be estimated in a fully consistent way (Sect.~\ref{section:AS}).
To investigate the influence of the applied halo component on the results, the fitting is redone with the original halo mass distribution being replaced by a truncated, flat rotation curve model according to \citet{wilkinson_evans} (hereafter denoted Model~II, Sect.~\ref{section:TF}) and a \citet{NFW} dark matter halo (hereafter denoted Model~III, Sect.~\ref{section:NFW}).
 
The spatial origin of the hypervelocity star HE~0437-5439 \citep{Edelmann_etal} is investigated as a first application example (Sect.~\ref{HE04375439}). Finally, conclusions are drawn (Sect.~\ref{section:conclusions}).
\section{Observations and fitting}\label{ObservationsandFitting}
In this section, all the details of the fitting process are outlined, i.e., what assumptions are made, what observations are taken into consideration, and how the fitting actually is done.
\subsection{Introductory remarks}
All gravitational potentials $\Phi(r,z)$ considered here are the sum of a central spherical bulge component $\Phi_{\mathrm{b}}(R)$, an axisymmetric disk $\Phi_{\mathrm{d}}(r,z)$, and a massive spherical dark matter halo $\Phi_{\mathrm{h}}(R)$:
\begin{equation}
\Phi(r,z) = \Phi_{\mathrm{b}}(R(r,z)) + \Phi_{\mathrm{d}}(r,z) + \Phi_{\mathrm{h}}(R(r,z))\, . \;
\label{Phi}
\end{equation}
Here, $(r,\phi,z)$ are cylindrical coordinates and $R=R(r,z)=\sqrt{r^2+z^2}$ is the spherical radius.

The corresponding density can be derived by virtue of Poisson's equation:
\begin{equation}
4 \pi G \rho(r,z) = \nabla^2 \Phi(r,z)\, . \;
\label{poisson}
\end{equation}

For the sake of simplicity, all potentials are independent of time. Consequently, rotating, nonaxisymmetric features, such as the Galactic bar or the spiral arms, are not implemented here, because they require that more complex models \citep[see, e.g.,][]{pichardo_etal, pichardo_etal_bar} are considered. To account for the effects of bar and spiral arms, additional systematic uncertainties are added to the observational data where necessary, see for instance Sect.~\ref{RotationCurve}.

Sticking to the convention of \citet{Allen_Santillan}, throughout this paper, the gravitational potentials are expressed in units of $100\,\mathrm{km}^2\,\mathrm{s}^{-2}$, lengths in $\mathrm{kpc}$, and masses in galactic mass units $\mathrm{M}_{\mathrm{gal}} = 100\times 1000^2\,\mathrm{kpc}\,[\mathrm{m}]/G\,[\mathrm{SI}]\,\mathrm{M}_{\sun} \approx 2.325\times 10^7\,\mathrm{M}_{\sun}$, yielding a gravitational constant $G$ of unity.
\subsection{Observational constraints}
Similar to \citet{Dehnen_Binney}, observations are divided into eight different groups that are introduced and discussed in the following.
\subsubsection{Solar kinematics}
Analyzing kinematic observations made in the celestial reference frame in a Galactocentric picture requires a coordinate transformation from one system to the other. Since this transformation obviously depends on the Sun's position and velocity, the interpretation of any data is affected by the values chosen for these parameters. To set up the coordinate transformation, we take J2000.0 coordinates for the Galactic center (GC) $\alpha_{\mathrm{GC}} = 17^{\mathrm{h}}45^{\mathrm{m}}37\fs224$, $\delta_{\mathrm{GC}} = -28\degr56\arcmin10\farcs23$ and north Galactic pole (NGP) $\alpha_{\mathrm{NGP}} = 12^{\mathrm{h}}51^{\mathrm{m}}26\fs282$, $\delta_{\mathrm{NGP}} = +27\degr07\arcmin42\farcs01$, as mentioned in the appendix of \citet{Reid_Brunthaler}. By monitoring stellar orbits around the central supermassive black hole Sgr~A*, \citet{Gillessen_etal} restricted the distance $r_{\sun}$ from the Sun to the GC to the range 
\begin{equation}
r_{\sun} = 8.33 \pm 0.35\,\mathrm{kpc} \, , \;
\label{SunGCDist}
\end{equation} 
which is the first observational constraint.

The Sun's velocity is a superposition of the in-plane circular motion $\varv_0$ of its local standard of rest (LSR) around the GC and its peculiar motion $\vec{\varv}_{\sun} = \left(U,V,W\right)_{\sun}$ relative to the LSR. Hereby, $U$ is the component towards the GC, $V$ in direction of Galactic rotation, and $W$ perpendicular to the Galactic plane. Motivated by the \citeauthor{McMillan_Binney} \citeyearpar{McMillan_Binney} discussion of a systematic offset in the $V$-component of maser motions versus a nonstandard value of $V_{\sun}$, the $\left(U,V,W\right)_{\sun}$ values determined by \citet{Schonrich_etal} are in this case preferable to others since they allow the use of maser sources to constrain the Galactic rotation curve (see Sect.~\ref{RotationCurve}) without the necessity of modeling an additional net peculiar motion. Thus, $\vec{\varv}_{\sun} = (11.1_{-0.75}^{+0.69}, 12.24_{-0.47}^{+0.47}, 7.25_{-0.36}^{+0.37}) \pm (1,2,0.5)\, \mathrm{km}\,\mathrm{s}^{-1}$ is used here.

Following the argument in \citet{McMillan}, the proper motion of Sgr~A* along Galactic longitude $l$, $\mu_{\mathrm{SgrA^*}} = -6.379 \pm 0.026\,\mathrm{mas}\,\mathrm{yr}^{-1}$ \citep{Reid_Brunthaler} is utilized as an indicator for $\varv_0$ instead of Oort's constants $A$ and $B$. With a spatial motion of Sgr~A* of $0 \pm 1\,\mathrm{km}\,\mathrm{s}^{-1}$ \citep{McMillan_Binney}, this proper motion can be assumed to be solely caused by the solar motion around the GC yielding as second observational constraint:
\begin{equation}
\mu_{\mathrm{SgrA^*}} = -\frac{1}{r_{\sun}} \left( V_{\sun} + \varv_0 \right)= -\frac{1}{r_{\sun}} \left( V_{\sun} + \varv_{\mathrm{c}}(r_{\sun}) \right)\, . \;
\label{musgra}
\end{equation}
The circular velocity $\varv_{\mathrm{c}}(r)$ is hereby linked to the potential via
\begin{equation}
\varv_{\mathrm{c}}(r) = \left.\sqrt{\acute{r} {\frac{\mathrm{d}\Phi(\acute{r},0)}{\mathrm{d}\acute{r}}}}\right|_{\acute{r}=r}\, . \;
\label{vcirc}
\end{equation}
\subsubsection{Rotation curve}\label{RotationCurve}
The Galactic rotation curve deduced from terminal velocities in the inner and intermediate regions and from maser sources in the outer region yield the third, fourth, and fifth groups of observational constraints considered here.
\paragraph{Terminal velocities:}
the terminal velocity $\varv_{\mathrm{term}}(l)$ is the measured peak velocity of the interstellar gas along the line-of-sight at Galactic coordinates $b=0$ and $-90\degr < l < 90\degr$. Assuming a circularly rotating interstellar medium, $\varv_{\mathrm{term}}(l)$ can be computed from the circular velocity $\varv_{\mathrm{c}}$ at $r = r_{\sun} \sin(l)$ corrected for the observer's projected motion:
\begin{equation}
\varv_{\mathrm{term}}(l) = \varv_{\mathrm{c}}\left(r_{\sun}\sin\,l\right) - \left(\varv_{\mathrm{c}}(r_{\sun}) + V_{\sun}\right) \sin\,l - U_{\sun} \cos\,l\, . \;
\label{vterm}
\end{equation}
For this study, unprocessed terminal velocities are extracted from surveys in \ion{H}{i} by \citet{Burton_Gordon} and in \element[][]{CO} by \citet{Clemens}. Similar to \citet{Dehnen_Binney}, a constant uncertainty of $7\, \mathrm{km}\,\mathrm{s}^{-1}$ is added in quadrature to the stated uncertainties of the \citeauthor{Clemens} data to account for non circular motions and the inability of the model potential to reproduce, e.g., spiral arm features. Complementary to the \element[][]{CO} measurements, the \ion{H}{i} data probe the central region with $|\sin\,l| < 0.3$, which is additionally affected by distortions due to the bar. Therefore, a rather generous uncertainty of $15\, \mathrm{km}\,\mathrm{s}^{-1}$ is quadratically added to their measurement errors.
\paragraph{Maser observations:}
maser sources are associated with high-mass star-forming regions and are consequently good tracers of the kinematic properties of their surrounding interstellar gas. Using very long baseline interferometry techniques, their positions, parallaxes, and proper motions can be measured to very high accuracy even for distant sources giving the opportunity to sample a relatively wide range of Galactocentric radii. Together with heliocentric radial velocities from Doppler shifts, these data hence allow precise determination of the masers' three-dimensional motions. In contrast to \citet{masers4}, \citet{McMillan_Binney} show that it is fairly justified to adopt their corresponding circular velocity components as probes for the Galactic rotation curve when taking the \citeauthor{Schonrich_etal} values for the peculiar motion of the Sun. Following this argument, parallaxes, proper motions, and heliocentric radial velocities for 30 maser sources have been compiled from the literature \citep{masers1, masers2, masers3, masers4, masers5, masers6, masers7, masers8, masers9}. To account for the virial motion of the masers with respect to their high-mass star-forming regions, a constant uncertainty of $7\, \mathrm{km}\,\mathrm{s}^{-1}$ \citep[see][]{masers4} is added in quadrature to the measurement errors when computing circular velocities.
\subsubsection{Mass and surface densities}
The local dynamical mass density~$\rho_{\sun}$ and the surface density~$\Sigma_{1.1}$ are together the sixth group of observational constraints.  Their derivation is straightforward from the model:
\begin{equation}
\rho_{\sun} = \rho_{\mathrm{b}}(r_{\sun}) + \rho_{\mathrm{d}}(r_{\sun},0) + \rho_{\mathrm{h}}(r_{\sun})
\label{rho_sun}
\end{equation}
\begin{equation}
\Sigma_{1.1} = \int\limits_{-1.1\,\mathrm{kpc}}^{1.1\,\mathrm{kpc}} \left(\rho_{\mathrm{b}}(r_{\sun},z) + \rho_{\mathrm{d}}(r_{\sun},z) + \rho_{\mathrm{h}}(r_{\sun},z)\right) \mathrm{d}z\, . \;
\label{sigma_1.1}
\end{equation}
\citeauthor{Holmberg_Flynn_density} obtained $\rho_{\sun} = 0.102 \pm 0.010\,\mathrm{M}_{\sun}\,\mathrm{pc}^{-3}$ from {\sc Hipparcos} data on a volume-complete sample of A and F stars \citep{Holmberg_Flynn_density} and $\Sigma_{1.1} = 74 \pm 6\,\mathrm{M}_{\sun}\,\mathrm{pc}^{-2}$ from observations of K giant stars \citep{Holmberg_Flynn_surface}.
\subsubsection{Velocity dispersion in Baade's window}
The velocity dispersion of the bulge in Baade's window, $\sigma_{\mathrm{BW}}$, is implemented as the seventh constraint probing the inner most region of the Milky Way. Following the reasoning of \citet{Dehnen_Binney}, it is given as
\begin{equation}
\sigma_{\mathrm{BW}} = \sigma_{\mathrm{b}}(0.0175 r_{\sun}, -0.068 r_{\sun}) = 117 \pm 15\,\mathrm{km}\,\mathrm{s}^{-1} \, , \;
\label{sigma_BW}
\end{equation}
where $\sigma_{\mathrm{b}}$ is estimated from the model potential as
\begin{equation}
\sigma_{\mathrm{b}}^2(r,z) = \frac{1}{\rho_{\mathrm{b}}(r,z)} \int\limits_{z}^{\infty}\rho_{\mathrm{b}}(r,\acute{z}) \frac{\partial \Phi(r,\acute{z})}{\partial \acute{z}} \mathrm{d}\acute{z} \, . \;
\label{sigma_b}
\end{equation}
\subsubsection{The extreme halo star J1539+0239}
The most stringent constraints on the halo mass result from the kinematically most extreme objects. Usually, the very distant satellite galaxy Leo~I is considered to be the most important tracer of the halo mass. Its proper motion has recently been measured with the Hubble Space Telescope (HST) for the first time \citep{leoi} and allowed to determine a limit on the halo mass \citep{leoimass}. However, close-by halo stars may provide even more compelling results. The kinematically most extreme halo star is the blue horizontal-branch star J1539+0239 for which \citet{J15390239} found a Galactic rest-frame velocity of about $694_{-221}^{+300}\,\mathrm{km}\,\mathrm{s}^{-1}$, making it the fastest halo object known so far. Unlike other hypervelocity stars, however, this object is currently approaching the Galactic disk with its pericentric passage of the GC still to occur. As discussed in the next paragraph, this indicates that J1539+0239 is gravitationally bound to the Milky Way, yielding a so far unexploited and --~due to the star's immense space motion~-- significant constraint on the Galaxy's total mass. We show in Sects.~\ref{section:AS_Result}, \ref{section:TF_Result}, and \ref{section:NFW_Result} that the limits on the halo mass derived from the kinematics of this star alone are even more stringent than those from Leo~I, so we add J1539+0239 to our set of constraints for the potentials and use Leo~I as a consistency check. Our approach, which is based on the six-dimensional phase space information of the most extreme objects, is complementary to statistical studies of several thousand stars
\citep[see, e.g.,][]{xue_etal, Gnedin_etal} that only make use of line-of-sight velocities.

To justify the assumption that J1539+0239 is bound to the Galaxy, consider the opposite, i.e., a trajectory that passes by the Milky Way only once, never to come back again. Such an orbit could be explained either by a non-Galactic origin, e.g., as a hypervelocity star ejected from another galaxy, or by an extreme dynamical event, e.g., a very close multibody encounter in a globular cluster or satellite galaxy, ejecting a star from its halo environment in a direction pointing not too far away from the GC. Although both scenarios cannot be ruled out by stellar age or space motion arguments, they are unlikely since they require very special circumstances, e.g., an extraordinarily high ejection velocity, a specific ejection direction, or the ejection event occurring at a certain time. Most of these restrictions are weaker or do not apply at all in the case of a bound orbit. For instance, the peculiar motion of J1539+0239 could still be the result of an ejection event. But because the accessible phase space is finite now, there is a non-zero probability of finding a trajectory starting from any initial condition (with matching conserved quantities) sooner or later in a phase-space state very close to J1539+0239, allowing for countless more ejection locations, directions and times. In particular, the GC no longer can be excluded as the star's spatial origin, which opens up the possibility of a dynamical interaction with the central supermassive black hole \citep{Hills}. This mechanism is known to be most powerful in terms of acceleration and is the best candidate for explaining extreme velocity objects. Alternatively, as shown by \citet{Abadi_etal}, the latter may result from the tidal disruption of a dwarf galaxy, a non ejection scenario that is also able to produce bound (but not unbound), high-velocity stars that approach the Galactic disk.

The assumption that J1539+0239 is gravitationally bound to the Milky Way is the eighth and last observational constraint. With $E_{*}$ denoting the star's total energy, i.e., the sum of kinetic and potential energy, per unit mass, the standard approach to incorporating this is to require $E_{*} \le 0$. But the Milky Way is not an isolated system, and beyond some distance, the influence of other objects such as Andromeda is no longer negligible. Since those effects are not included in the presented models, a more rigorous approach is chosen here by defining a bound state as
\begin{equation}
E_{*} \le \Phi(0,200\,\mathrm{kpc}) \, , \;
\label{boundness_condition}
\end{equation}
theoretically limiting orbits to $R \sim 200\,\mathrm{kpc}$. However, the actual implementation in the least-squares fit is
\begin{subnumcases}{\chi_{*} =\label{boundness_condition_chi}}
\frac{E_{*} - \Phi(0,200\,\mathrm{kpc})}{\Delta E_{*}} & if $E_{*} - \Phi(0,200\,\mathrm{kpc}) > 0$,\\
0 & otherwise.
\end{subnumcases}
Hereby, $\Delta E_{*}$ is the error in $E_{*}$ resulting from uncertainties in determining J1539+0239's distance, radial velocity, and proper motions. Since $\Delta E_{*}$ is large (for instance $E_{*}/\Delta  E_{*} \sim 0.82$ in the original \citeauthor{Allen_Santillan} potential), violations of Eq.~(\ref{boundness_condition}) are generously admitted. For the same reason, the particular choice of $200\,\mathrm{kpc}$ is not decisive as for instance $-\Phi(0,200\,\mathrm{kpc})/\Delta  E_{*} \sim 0.16$ while $-\Phi(0,150\,\mathrm{kpc})/\Delta  E_{*} \sim 0.21$ and $-\Phi(0,250\,\mathrm{kpc})/\Delta  E_{*} \sim 0.13$ (again in the original \citeauthor{Allen_Santillan} potential).
\subsection{Fitting process}
The model parameters (see Sects.~\ref{section:AS}--\ref{section:NFW}) are simultaneously determined via a $\chi^2$-minimization procedure using the Interactive Spectral Interpretation System \citep[ISIS,][]{isis}. Observations of the rotation curve, as well as of J1539+0239, are always taken in raw form from the literature, i.e., as measured in celestial coordinates, in order to exclude inconsistencies during the transformation to the Galactic reference frame originating in different values for the LSR and solar motion. Moreover, any given uncertainty found in the literature is considered here, typically via Gaussian statistics. For instance, to compare the maser measurements to model predictions, their positions and velocities are converted to Galactic coordinates, allowing  computation of rotational velocities $\varv_{\mathrm{c}}$ and Galactrocentric distances $r$. The corresponding uncertainties in the observational data are then accounted for using Gaussian error propagation. Additionally, the (admittedly small) uncertainties stemming from $\vec{\varv}_{\sun}$ are propagated in the same way. As the resulting data points in the ($r,\varv_{\mathrm{c}}$)-plane have errors in both directions ($\Delta r$, $\Delta \varv_{\mathrm{c}}$), it is not obvious at which radius to compare the model to observation, i.e., how to compute the $\chi^2$ without neglecting $\Delta r$. The solution chosen here is to convert $\Delta r$ to an error in $\varv_{\mathrm{c}}$ by estimating the effects of it on the model rotation curve according to $(\Delta \varv_{\mathrm{c}})_{\mathrm{model}} = (\mathrm{d}\varv_{\mathrm{c}}/\mathrm{d}r)_{\mathrm{model}} \Delta r$ and adding it in quadrature to $\Delta \varv_{\mathrm{c}}$.

Very much attention is paid to take all sources of error into consideration in order to assign the correct weighting to each observational constraint in the total $\chi^2$. Unfortunately, systematic effects such as the missing modeling of spiral arms are difficult to quantify and are not sufficiently dealt with by just adding a constant error. Therefore, an unweighted fitting routine, e.g., assuming that a single maser point weighs the same as the proper motion of Sgr~A* (see Eq.~(\ref{musgra})), does not yield a satisfactory result because the fit would be dominated by the large number of data points used while fitting the rotation curve. To account for this, each contribution to the total $\chi^2$ of the above-mentioned eight groups of observational constraints is divided by the number of data points $N$ within the group:
\begin{equation}
\chi_{\mathrm{w}}^2 = \sum\limits_{i=1}^{8} \left( \frac{1}{N_i} \sum\limits_{j=1}^{N_i} \chi_{ij}^2 \right) \, . \;
\label{chi}
\end{equation}
The model parameters given in Tables~\ref{table:results_AS}--\ref{table:results_NFW} are those that minimize this weighted $\chi^2$.
\section{Model~I}\label{section:AS}
Model~I is a revision of the \citet{Allen_Santillan} potential.
\subsection{Components and characteristics}
\subsubsection{Bulge and disk}\label{section:bulge_disk}
The potential forms $\Phi_{\mathrm{b}}(R)$ and $\Phi_{\mathrm{d}}(r,z)$ of bulge and disk are those proposed by \citet{Miyamoto_Nagai}:
\begin{equation}
\Phi_{\mathrm{b}}(R) = -\frac{M_{\mathrm{b}}}{\sqrt{R^2+{b_{\mathrm{b}}}^2}} 
\label{Phi_b}
\end{equation}
\begin{equation}
\Phi_{\mathrm{d}}(r,z) = -\frac{M_{\mathrm{d}}}{\sqrt{r^2 + \left({a_{\mathrm{d}}} + \sqrt{z^2 +{b_{\mathrm{d}}}^2}\right)^2}}\, . \;
\label{Phi_d}
\end{equation}
The parameters $M_{\mathrm{b}}$/$M_{\mathrm{d}}$ determine the contribution of their components to the total potential, and $b_{\mathrm{b}}$/$a_{\mathrm{d}}$/$b_{\mathrm{d}}$ are scale lengths.

The related densities $\rho_{\mathrm{b}}$ and $\rho_{\mathrm{d}}$ are (see Eq.~(\ref{poisson}))
\begin{equation}
\rho_{\mathrm{b}}(R) = \frac{3 {b_{\mathrm{b}}}^2 M_{\mathrm{b}}}{4 \pi ({R^2+{b_{\mathrm{b}}}^2})^{5/2}} 
\label{rho_b}
\end{equation}
\begin{equation}
\rho_{\mathrm{d}}(r,z) = \frac{{b_{\mathrm{d}}}^2 M_{\mathrm{d}}}{4 \pi} \frac{{a_{\mathrm{d}}} r^2 + \left({a_{\mathrm{d}}} + 3 \sqrt{z^2+{b_{\mathrm{d}}}^2}\right)\left({a_{\mathrm{d}}} + \sqrt{z^2+{b_{\mathrm{d}}}^2}\right)^2}{\left(z^2+{b_{\mathrm{d}}}^2\right)^{3/2} \left(r^2+\left({a_{\mathrm{d}}}+\sqrt{z^2+{b_{\mathrm{d}}}^2}\right)^2\right)^{5/2}}\, . \;
\label{rho_d}
\end{equation}
Integrating these densities over the entire volume to obtain the total masses $m_{\mathrm{b}}$ and $m_{\mathrm{d}}$ of bulge and disk gives the expected identities $m_{\mathrm{b}} = M_{\mathrm{b}}$ and $m_{\mathrm{d}} = M_{\mathrm{d}}$.
\subsubsection{The dark matter halo}
Pursuing mathematical simplicity, the halo potential $\Phi_{\mathrm{h}}(R)$ used here differs slightly from the original one by \citeauthor{Allen_Santillan}. We assume the halo mass $m_{\mathrm{h}}$ inside a sphere of radius $R$ is given by \citep[see][]{Allen_Martos}
%\begin{eqnarray}
%m_{\mathrm{h}}(<R) & = & \frac{M_{\mathrm{h}} \left(\frac{R}{a_{\mathrm{h}}}\right)^\gamma }{1 + \left(\frac{R}{a_{\mathrm{h}}}\right)^{\gamma-1}}\, , \; R < \Lambda\nonumber \\
%                  & = & \frac{M_{\mathrm{h}} \left(\frac{\Lambda}{a_{\mathrm{h}}}\right)^\gamma }{1 + \left(\frac{\Lambda}{a_{\mathrm{h}}}\right)^{\gamma-1}} = \mathrm{constant}\, , \; R \gid \Lambda\, . \;
%\label{m_h}
%\end{eqnarray}
\begin{subnumcases}{m_{\mathrm{h}}(<R) =\label{m_h}}
\frac{M_{\mathrm{h}} \left(\frac{R}{a_{\mathrm{h}}}\right)^\gamma }{1 + \left(\frac{R}{a_{\mathrm{h}}}\right)^{\gamma-1}} & if $R < \Lambda$,\\
\frac{M_{\mathrm{h}} \left(\frac{\Lambda}{a_{\mathrm{h}}}\right)^\gamma }{1 + \left(\frac{\Lambda}{a_{\mathrm{h}}}\right)^{\gamma-1}} = \mathrm{constant,} & otherwise.
\end{subnumcases}
Here, $M_{\mathrm{h}}$ is again a weighting factor and $a_{\mathrm{h}}$ a scale length. Clearly, this specific form is chosen to have the asymptotic behavior $m_{\mathrm{h}}(<R) \propto R$ for large $R$ motivated by the observed flat rotation curve. To avoid an unphysical, infinite halo mass, a cutoff parameter $\Lambda$ is incorporated as well. In principle, the exponent $\gamma$ is a free parameter, too. However, the shortage of halo constraints means the ability of the model to reproduce the observations is not attenuated by setting $\gamma = 2$, reducing the expression's complexity (and avoiding singularities at the origin in the equations of motion for $\gamma < 2$).

The expression $m_{\mathrm{h}}(<R)$ is the volume integral of the halo density $\rho_{\mathrm{h}}$ over the sphere of radius $R$. Assuming a spherically symmetric halo density $\rho_{\mathrm{h}} = \rho_{\mathrm{h}}(R)$ and using Poisson's equation (Eq.~(\ref{poisson})), a relation between $m_{\mathrm{h}}(<R)$ and its respective potential can be derived:
\begin{eqnarray}
m_{\mathrm{h}}(<R) & = & 4 \pi \int\limits_{0}^{R} \acute{R}^2 \rho_{\mathrm{h}}(\acute{R}) \mathrm{d}\acute{R} = \int\limits_{0}^{R} \acute{R}^2 \nabla^2 \Phi_{\mathrm{h}}(\acute{R}) \mathrm{d}\acute{R} \nonumber \\ 
                  & = & \int\limits_{0}^{R} \acute{R}^2 \frac{1}{\acute{R}^2} \frac{\mathrm{d}}{\mathrm{d}\acute{R}} \left(\acute{R}^2 \frac{\mathrm{d}}{\mathrm{d}\acute{R}} \Phi_{\mathrm{h}}(\acute{R})\right) \mathrm{d}\acute{R} \nonumber \\ 
                  & = & R^2 \frac{\mathrm{d}}{\mathrm{d}R} \Phi_{\mathrm{h}}(R)\, . \;
\end{eqnarray}
Inserting Eq.~(\ref{m_h}) and directly integrating the result accounting for the boundary condition $\Phi_{\mathrm{h}}(\infty) = 0$ yields
\[
\Phi_{\mathrm{h}}(R) = \int\limits_{\infty}^{R} \frac{m_{\mathrm{h}}(\acute{R})}{\acute{R}^2} \mathrm{d}\acute{R} =  %\nonumber \\
\]
\begin{subnumcases}{}
\frac{M_{\mathrm{h}} }{a_{\mathrm{h}}} \left( \frac{1}{(\gamma-1)} \ln\left(\frac{1+\left(\frac{R}{a_{\mathrm{h}}}\right)^{\gamma-1}}{1+\left(\frac{\Lambda}{a_{\mathrm{h}}}\right)^{\gamma-1}} \right) -  \frac{\left(\frac{\Lambda}{a_{\mathrm{h}}}\right)^{\gamma-1}}{1+\left(\frac{\Lambda}{a_{\mathrm{h}}}\right)^{\gamma-1}} \right) & \hspace{-2ex} if $R < \Lambda$,\\
- \frac{M_{\mathrm{h}}}{R} \frac{\left(\frac{\Lambda}{a_{\mathrm{h}}}\right)^{\gamma}}{1+\left(\frac{\Lambda}{a_{\mathrm{h}}}\right)^{\gamma-1}}, & \hspace{-2ex} otherwise.
\end{subnumcases}
For $\Lambda = 100\,\mathrm{kpc}$ and $\gamma = 2.02$, this is equivalent to the expression given in \citet{Allen_Santillan} obtained via integration by parts. The advantage of the above representation is that the dependent variable $R$ appears only once and linearly in the argument of the logarithm (since $\gamma = 2$).

The corresponding density $\rho_{\mathrm{h}}$ is
\begin{subnumcases}{\rho_{\mathrm{h}}(R) = }
\frac{M_{\mathrm{h}}}{4 \pi a_{\mathrm{h}}} \frac{\left(\frac{R}{a_{\mathrm{h}}}\right)^{\gamma-1} \left( \left(\frac{R}{a_{\mathrm{h}}}\right)^{\gamma-1} + \gamma \right)}{R^2 \left(1+\left(\frac{R}{a_{\mathrm{h}}}\right)^{\gamma-1} \right)^2} & if $R < \Lambda$,\\
0 & otherwise.
\end{subnumcases}
\subsection{Results}\label{section:AS_Result}
\begin{table*}
\renewcommand{\arraystretch}{1.2}
\caption{\label{table:results_AS}Properties of the best-fitting Model~I.}
\centering
\begin{tabular}{lc|ccc|cc}
\hline\hline
Parameter & Value & Constraint & \multicolumn{2}{c}{Value} \vline& Derived Quantity & Value\\
\cline{4-5}
          &       &            & Observation & Model       &                  &      \\
\hline
$r_{\sun}\,(\mathrm{kpc})$ & $8.40\pm0.08$ & $r_{\sun}\,(\mathrm{kpc})$ & $8.33 \pm 0.35$ & $8.40$ & $\varv_0\,(\mathrm{km}\,\mathrm{s}^{-1})$ & $242.0$\\
$M_{\mathrm{b}}\,(\mathrm{M}_{\mathrm{gal}})$ & $409\pm63$ & $\mu_{\mathrm{SgrA^*}}\,(\mathrm{mas}\,\mathrm{yr}^{-1})$ & $-6.379 \pm 0.026$ & $-6.384$ & $m_{\mathrm{b}}\,(10^{9}\mathrm{M}_{\sun})$ & $9.5 \pm 1.5$ \\
$M_{\mathrm{d}}\,(\mathrm{M}_{\mathrm{gal}})$ & $2856^{+376}_{-202}$ & Terminal velocities from \ion{H}{i} & See Fig.~\ref{fig:rotation_curve_AS}  & See Fig.~\ref{fig:rotation_curve_AS} & $m_{\mathrm{d}}\,(10^{10}\mathrm{M}_{\sun})$ & $6.6^{+0.9}_{-0.5}$\\
$M_{\mathrm{h}}\,(\mathrm{M}_{\mathrm{gal}})$ & $1018^{+27933}_{-603}$\tablefootmark{a} & Terminal velocities from \element[][]{CO} & See Fig.~\ref{fig:rotation_curve_AS}  & See Fig.~\ref{fig:rotation_curve_AS} & $m_{\mathrm{h}}\,(10^{12}\mathrm{M}_{\sun})$ & $1.8^{+2.4}_{-0.8}$\\
$b_{\mathrm{b}}\,(\mathrm{kpc})$ & $0.23\pm0.03$ & Circular velocities from masers & See Fig.~\ref{fig:rotation_curve_AS}  & See Fig.~\ref{fig:rotation_curve_AS} & $M_{R<50\,\mathrm{kpc}}\,(10^{12}\mathrm{M}_{\sun})$ & $0.51^{+0.33}_{-0.04}$\\
$a_{\mathrm{d}}\,(\mathrm{kpc})$ & $4.22^{+0.53}_{-0.99}$ & $\rho_{\sun}\,(\mathrm{M}_{\sun}\,\mathrm{pc}^{-3})$, $\Sigma_{1.1}\,(\mathrm{M}_{\sun}\,\mathrm{pc}^{-2})$ & $0.102 \pm 0.010$, $74 \pm 6$ & $0.102$, $74$ & $M_{R<100\,\mathrm{kpc}}\,(10^{12}\mathrm{M}_{\sun})$ & $0.97^{+0.96}_{-0.09}$\\
$b_{\mathrm{d}}\,(\mathrm{kpc})$ & $0.292^{+0.020}_{-0.025}$ & $\sigma_{\mathrm{BW}}\,(\mathrm{km}\,\mathrm{s}^{-1})$ & $117 \pm 15$ & $120$ & $M_{R<200\,\mathrm{kpc}}\,(10^{12}\mathrm{M}_{\sun})$ & $1.9^{+2.4}_{-0.8}$\\
$a_{\mathrm{h}}\,(\mathrm{kpc})$ & $2.562^{+25.963}_{-1.419}$\tablefootmark{a} & $\chi_{*}$ & $\le 0$ & $0.66$ & $\varv_{\mathrm{esc},\sun}\,(\mathrm{km}\,\mathrm{s}^{-1})$ & $616.4$\\
$\Lambda\,(\mathrm{kpc})$ & $200^{+0}_{-83}$\tablefootmark{b} &  &  &  & $A\,(\mathrm{km}\,\mathrm{s}^{-1}\,\mathrm{kpc}^{-1})$ & $15.06$\\
$\gamma$ & $2$ (fixed) & & & & $B\,(\mathrm{km}\,\mathrm{s}^{-1}\,\mathrm{kpc}^{-1})$ & $-13.74$\\ 
\hline
\end{tabular}
\tablefoot{
The quoted uncertainties for the model parameters and masses are 90\%-confidence limits: After normalizing the weighted $\chi_{\mathrm{w}}^2$ via multiplication with a factor yielding $\chi_{\mathrm{w}}^2 / \mathrm{d.o.f.} = 1$ at the minimum $\chi_{\mathrm{w}}^2$, 90\%-confidence intervals are calculated from this normalized $\chi^2$ statistics and the condition $\Delta \chi^2 = 2.71$. \tablefoottext{a}{The large uncertainties are due to a strong correlation between $M_{\mathrm{h}}$ and $a_{\mathrm{h}}$, see Fig.~\ref{fig:ah_mh_AS}.} \tablefoottext{b}{Motivated by cosmological studies, the restriction $\Lambda \le 200\,\mathrm{kpc}$ is imposed.}
}
\end{table*}
\begin{figure*}
\centering
\includegraphics[width=\textwidth]{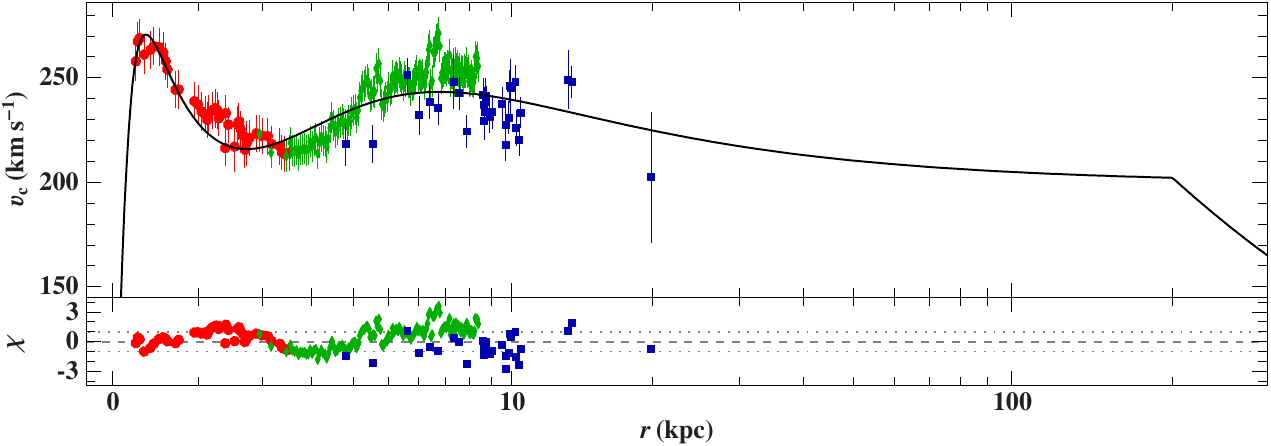}
\caption{Comparison of the best-fitting model rotation curve (solid black line) with terminal velocities from surveys in \ion{H}{i} (red circles) and in \element[][]{CO} (green diamonds), as well as maser observations (blue squares) for Model~I.}
\label{fig:rotation_curve_AS}
\end{figure*}
\begin{figure*}
\centering
\includegraphics[width=\textwidth]{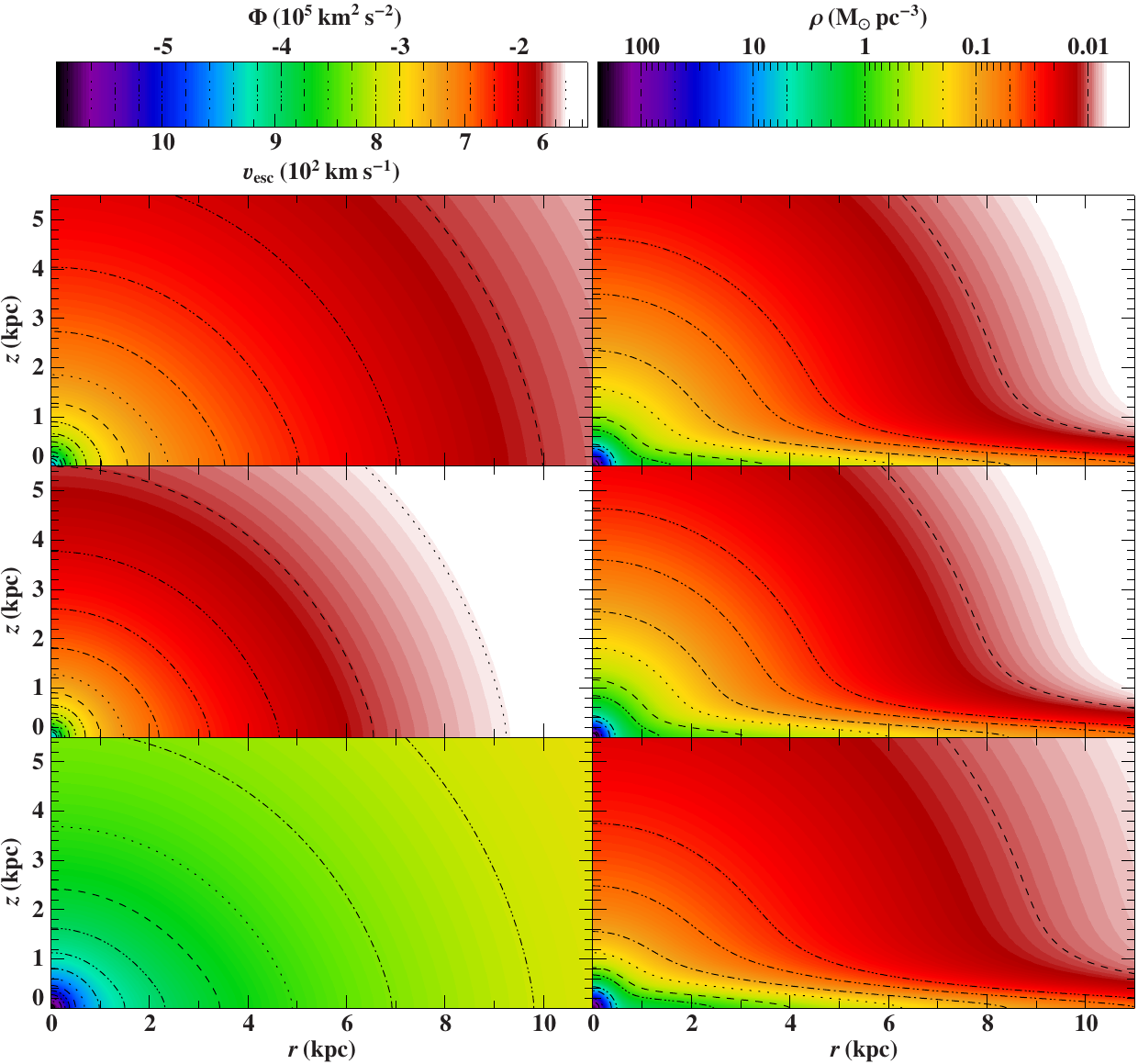}
%\resizebox{\hsize}{!}{\includegraphics{equi_phi_rho.pdf}}
\caption{Gravitational potential $\Phi(r,z)$, formal escape velocity $\varv_{\mathrm{esc}}(r,z) = \sqrt{-2 \Phi}$, and total mass density $\rho(r,z)$ for the best-fit parameters of Model~I (\textit{upper panel}), Model~II (\textit{middle panel}), and Model~III (\textit{lower panel}). Contours are indicated by dashed lines.}
\label{fig:phi_rho}
\end{figure*}
\begin{figure}
\centering
\includegraphics[width=0.48\textwidth]{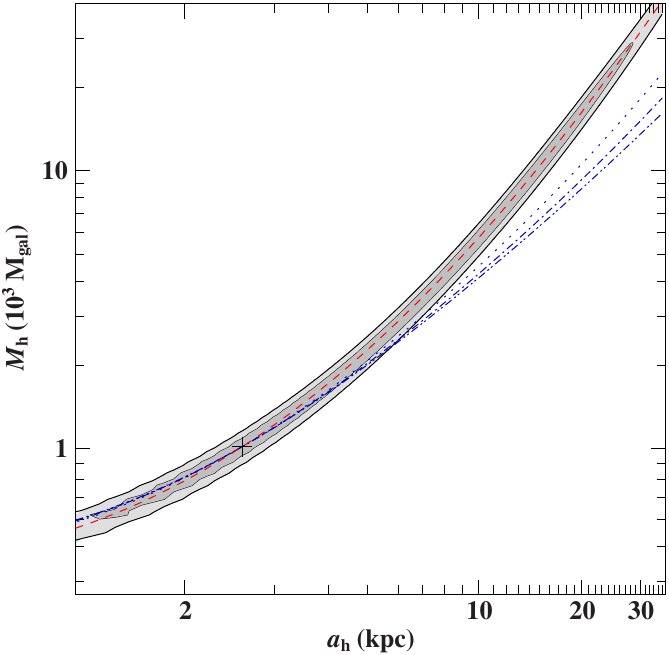}
\caption{Visualization of the correlation between $M_{\mathrm{h}}$ and $a_{\mathrm{h}}$ in Model~I: The single parameter 90\%-confidence region defined by $\Delta \chi^2 \le 2.71$ and corresponding to a 74\%-joint-confidence region is black-rimmed and shaded in dark gray, while the $3\sigma$ region ($\Delta \chi^2 \le 9$, 99\%-joint-confidence region) is the total black-rimmed, gray-shaded area. The cross marks the location of the best fit. The four curves define the loci of constant halo mass inside a sphere of radius $14.1\,\mathrm{kpc}$ (red dashed), $50\,\mathrm{kpc}$ (blue dotted), $100\,\mathrm{kpc}$ (blue dashed-dotted), and $200\,\mathrm{kpc}$ (blue dashed-dotted-dotted).}
\label{fig:ah_mh_AS}
\end{figure}
The properties of the best-fitting Model~I are visualized in Figs.~\ref{fig:rotation_curve_AS} and \ref{fig:phi_rho} and summarized in Table~\ref{table:results_AS}. Given the model's simplicity, the overall agreement with the imposed constraints is very good. In particular, $r_{\sun}$, $\mu_{\mathrm{SgrA^*}}$, $\rho_{\sun}$, $\Sigma_{1.1}$, and $\sigma_{\mathrm{BW}}$ are almost perfectly reproduced. The rotation curve, on the other hand, reveals systematic discrepancies between model and observation originating mainly in the inadequate treatment of the Galactic bar and the spiral arms. Moreover, most of the maser sources still seem to lag behind the Galactic rotation curve. Despite these shortcomings, the unweighted $\chi^2$ per degree of freedom (d.o.f.) is about $1.5$, thereby confirming the good match.

To see how well the individual parameters are confined by the observational constraints, 90\%-confidence limits were computed from the weighted $\chi_{\mathrm{w}}^2$ statistics after multiplying it with a factor yielding $\chi_{\mathrm{w}}^2 / \mathrm{d.o.f.} = 1$ at the best fit, i.e., at the minimum $\chi_{\mathrm{w}}^2$. The resulting intervals are given in Table~\ref{table:results_AS}, too. As illustrated in Fig.~\ref{fig:ah_mh_AS}, there is a strong correlation between $M_{\mathrm{h}}$ and $a_{\mathrm{h}}$ that explains the large uncertainties of these two parameters. This degeneracy is a consequence of the missing constraints on the halo mass distribution. The available observations are only able to constrain the halo mass inside a sphere of radius $14.1\,\mathrm{kpc}$ (see Fig.~\ref{fig:ah_mh_AS}), which, however, can be reproduced by numerous different combinations of $M_{\mathrm{h}}$-$a_{\mathrm{h}}$ pairs. A parametrization of the halo using only one independent variable would thus be sufficient to obtain a good fit to the data. From Fig.~\ref{fig:ah_mh_AS}, it is also clear that the degeneracy in the two halo parameters does not follow curves of constant halo mass inside $50$, $100$, or $200\,\mathrm{kpc}$, which  explains the relatively large uncertainties of the total masses inside these radii as listed in Table~\ref{table:results_AS}.

\citet{J15390239} were the first to exploit the star J1539+0239 as a probe for the mass of the Galaxy and its main contributor, the dark matter halo, and they derived $m_{\mathrm{h}} \ge 1.7^{+2.3}_{-1.1}\times10^{12}\,\mathrm{M}_{\sun}$ based on the same model potential as discussed here. Their result, however, was derived solely from the condition that J1539+0239 is bound to the Milky Way while neglecting any other observational constraint that keeps the mass low. In contrast, the mass estimates presented in Table~\ref{table:results_AS} are fully consistent with all the observational data mentioned before. 

The best-fit parameters yield a total mass (with 90\%-confidence limits computed analogously to the model parameters' uncertainties) of $M_{\mathrm{total}} = 1.9^{+2.4}_{-0.8} \times10^{12}\,\mathrm{M}_{\sun} \approx m_{\mathrm{h}}$. This result is a modest revision of the \citet{J15390239} value, and it confirms their conclusion that a total Galactic mass of about $1 \times 10^{12}\,\mathrm{M}_{\sun}$ such as found by \citet{xue_etal} possibly underestimates the true value. From Eq.~(\ref{m_h}) and the fact that $\Lambda = 200\,\mathrm{kpc}$ at the best fit, it is obvious that $M_{\mathrm{total}}$ is affected by the restriction $\Lambda \le 200\,\mathrm{kpc}$. Although the latter is justified by cosmological simulations on galaxy formation, the specific choice for the upper boundary can of course be subject to discussion. But since even the mass corresponding to $\Lambda = 450\,\mathrm{kpc}$ lies within the 90\%-confidence interval of $M_{\mathrm{total}}$, this concern does not seriously change the result. To remove the $\Lambda$ dependence, consider the quantity $M_{R<50\,\mathrm{kpc}}$, i.e., the total mass within $50\,\mathrm{kpc}$. From the kinematic analysis of a combined sample of field horizontal-branch stars, globular clusters, and satellite galaxies, \citet{Sakamoto} have obtained $M_{R<50\,\mathrm{kpc}} = 5.4^{+0.1}_{-0.4}\times10^{11}\,\mathrm{M}_{\sun}$ agreeing with $M_{R<50\,\mathrm{kpc}} = 5.1^{+3.3}_{-0.4} \times10^{11}\,\mathrm{M}_{\sun}$ as found in this study. 

It is worth noting that J1539+0239 is not at all gravitationally bound to the Milky Way in the best-fitting model. This is possible because of the large uncertainty in its position and velocity determination resulting in a small contribution ($\chi_{*} = 0.66$) to the total $\chi_{\mathrm{w}}^2$. More precise kinematic information on the star would set tighter constraints on the Galactic potential, probably reducing the uncertainties in the mass estimates.

We also model the trajectory of the very distant satellite galaxy Leo~I based on the full six-dimensional phase space information \citep[position: $\alpha = 10^{\mathrm{h}}8^{\mathrm{m}}28\fs68$, $\delta = +12\degr18\arcmin19.7\arcsec$; distance: $256.7 \pm 13.3 \,\mathrm{kpc}$; radial velocity: $282.9 \pm 0.5\,\mathrm{km}\,\mathrm{s}^{-1}$ ; proper motion: $\mu_{\alpha}\cos\delta = -0.1140 \pm 0.0295\,\mathrm{mas} \,\mathrm{yr}^{-1}$, $\mu_{\delta} = -0.1256 \pm 0.0293\,\mathrm{mas} \,\mathrm{yr}^{-1}$,][]{leoi}. Leo~I is found to be formally bound to the Galaxy, showing that J1539+0239 is indeed the dominant constraint on the halo mass. 

Table~\ref{table:results_AS} also contains some derived quantities that are not directly fitted, such as $\varv_{\mathrm{esc},\sun}$ or $\varv_0 / r_{\sun}$. The escape velocity at the Sun's position is about $616\,\mathrm{km}\,\mathrm{s}^{-1}$, thereby lying very close to the 90\%-confidence limit $498\,\mathrm{km}\,\mathrm{s}^{-1} < \varv_{\mathrm{esc},\sun} < 608\,\mathrm{km}\,\mathrm{s}^{-1}$ of \citet{smith_etal}. The ratio of $\varv_0$ over $r_{\sun}$ is $28.8\,\mathrm{km}\,\mathrm{s}^{-1}\,\mathrm{kpc}^{-1}$ and thus somewhat less than the values found by \citet{masers4}, $30.3 \pm 0.9\,\mathrm{km}\,\mathrm{s}^{-1}\,\mathrm{kpc}^{-1}$, or \citet{McMillan_Binney}, $29.9-31.6\,\mathrm{km}\,\mathrm{s}^{-1}\,\mathrm{kpc}^{-1}$, from maser sources alone.

As pointed out before, the halo mass distribution is not well constrained by the observations, so different potential shapes of the halo can yield almost equal matches to the data but differ significantly in other properties, as for instance the total mass. To investigate this behavior, two other representations of the dark matter halo are considered in the following sections.
\section{Model~II}\label{section:TF}
In Model~II, the halo is replaced by the truncated, flat rotation curve model used in \citet{wilkinson_evans} and \citet{Sakamoto}.
\subsection{Components and characteristics}
While bulge and disk components have the same shape as in Model~I, the halo potential reads as \citep{wilkinson_evans}
\begin{equation}
\Phi_{\mathrm{h}}(R) = - \frac{M_{\mathrm{h}}}{a_{\mathrm{h}}}\ln\left(\frac{\sqrt{R^2+{a_{\mathrm{h}}}^2}+a_{\mathrm{h}}}{R}\right)\, . \;
\end{equation}
The resulting density
\begin{equation}
\rho_{\mathrm{h}}(R) = \frac{M_{\mathrm{h}}}{4 \pi} \frac{{a_{\mathrm{h}}}^2}{ R^2 \left(R^2 + {a_{\mathrm{h}}}^2 \right)^{3/2}}
\end{equation}
is cusped like $R^{-2}$ for $R \ll a_{\mathrm{h}}$ and falls off like $R^{-5}$ for $R \gg a_{\mathrm{h}}$. In this way, the corresponding rotation curve is flat in the inner regions, and the total halo mass $m_{\mathrm{h}} = M_{\mathrm{h}}$ is finite without invoking a cutoff parameter.
\subsection{Results}\label{section:TF_Result}
\begin{table*}
\renewcommand{\arraystretch}{1.2}
\caption{\label{table:results_TF}Properties of the best-fitting Model~II.}
\centering
\begin{tabular}{lc|ccc|cc}
\hline\hline
Parameter & Value & Constraint & \multicolumn{2}{c}{Value} \vline& Derived Quantity & Value\\
\cline{4-5}
          &       &            & Observation & Model       &                  &      \\
\hline
$r_{\sun}\,(\mathrm{kpc})$ & $8.35\pm0.08$ & $r_{\sun}\,(\mathrm{kpc})$ & $8.33 \pm 0.35$ & $8.35$ & $\varv_0\,(\mathrm{km}\,\mathrm{s}^{-1})$ & $240.4$\\
$M_{\mathrm{b}}\,(\mathrm{M}_{\mathrm{gal}})$ & $175\pm28$ & $\mu_{\mathrm{SgrA^*}}\,(\mathrm{mas}\,\mathrm{yr}^{-1})$ & $-6.379 \pm 0.026$ & $-6.383$ & $m_{\mathrm{b}}\,(10^{9}\mathrm{M}_{\sun})$ & $4.1 \pm 0.7$\\
$M_{\mathrm{d}}\,(\mathrm{M}_{\mathrm{gal}})$ & $2829 \pm 192$ & Terminal velocities from \ion{H}{i} & See Fig.~\ref{fig:rotation_curve_TF} & See Fig.~\ref{fig:rotation_curve_TF} & $m_{\mathrm{d}}\,(10^{10}\mathrm{M}_{\sun})$ & $6.6 \pm 0.5$\\
$M_{\mathrm{h}}\,(\mathrm{M}_{\mathrm{gal}})$ & $69\,725^{+5790}_{-20\,931}$\tablefootmark{a} & Terminal velocities from \element[][]{CO} & See Fig.~\ref{fig:rotation_curve_TF} & See Fig.~\ref{fig:rotation_curve_TF} & $m_{\mathrm{h}}\,(10^{12}\mathrm{M}_{\sun})$ & $1.6^{+0.2}_{-0.5}$\\
$b_{\mathrm{b}}\,(\mathrm{kpc})$ & $0.184\pm0.040$ & Circular velocities from masers & See Fig.~\ref{fig:rotation_curve_TF} & See Fig.~\ref{fig:rotation_curve_TF} & $M_{R<50\,\mathrm{kpc}}\,(10^{12}\mathrm{M}_{\sun})$ & $0.46 \pm 0.03$\\
$a_{\mathrm{d}}\,(\mathrm{kpc})$ & $4.85^{+0.41}_{-0.33}$ & $\rho_{\sun}\,(\mathrm{M}_{\sun}\,\mathrm{pc}^{-3})$, $\Sigma_{1.1}\,(\mathrm{M}_{\sun}\,\mathrm{pc}^{-2})$ & $0.102 \pm 0.010$, $74 \pm 6$ & $0.102$, $75$ & $M_{R<100\,\mathrm{kpc}}\,(10^{12}\mathrm{M}_{\sun})$ & $0.79^{+0.06}_{-0.08}$\\
$b_{\mathrm{d}}\,(\mathrm{kpc})$ & $0.305\pm0.020$ & $\sigma_{\mathrm{BW}}\,(\mathrm{km}\,\mathrm{s}^{-1})$ & $117 \pm 15$ & $116$ & $M_{R<200\,\mathrm{kpc}}\,(10^{12}\mathrm{M}_{\sun})$ & $1.2^{+0.1}_{-0.2}$\\
$a_{\mathrm{h}}\,(\mathrm{kpc})$ & $200^{+0}_{-60}$\tablefootmark{a,}\tablefootmark{b} & $\chi_{*}$ & $\le 0$ & $0.80$ & $\varv_{\mathrm{esc},\sun}\,(\mathrm{km}\,\mathrm{s}^{-1})$ & $575.9$\\
&  &  &  &  & $A\,(\mathrm{km}\,\mathrm{s}^{-1}\,\mathrm{kpc}^{-1})$ & $15.11$\\
&  &  &  &  & $B\,(\mathrm{km}\,\mathrm{s}^{-1}\,\mathrm{kpc}^{-1})$ & $-13.68$\\ 
\hline
\end{tabular}
\tablefoot{
The quoted uncertainties for the model parameters and masses are 90\%-confidence limits (see notes on Table~\ref{table:results_AS} for details). \tablefoottext{a}{The large uncertainties are due to a strong correlation between $M_{\mathrm{h}}$ and $a_{\mathrm{h}}$, see Fig.~\ref{fig:ah_mh_TF}.} \tablefoottext{b}{Motivated by cosmological studies, the restriction $a_{\mathrm{h}} \le 200\,\mathrm{kpc}$ is imposed.}
}
\end{table*}
\begin{figure*}
\centering
\includegraphics[width=\textwidth]{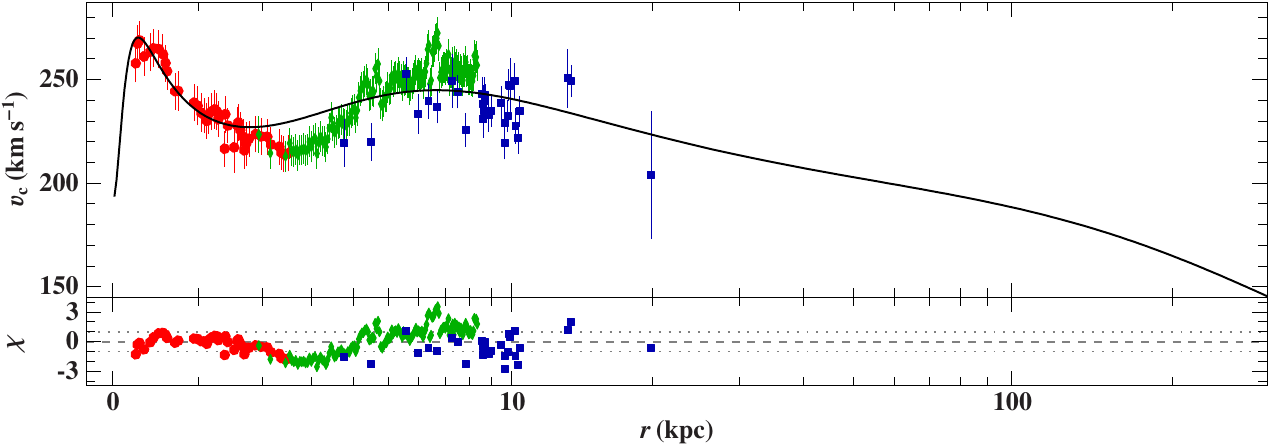}
\caption{Comparison of the best-fitting model rotation curve (solid black line) with terminal velocities from surveys in \ion{H}{i} (red circles) and in \element[][]{CO} (green diamonds), as well as maser observations (blue squares) for Model~II.}
\label{fig:rotation_curve_TF}
\end{figure*}
\begin{figure}
\centering
\includegraphics[width=0.48\textwidth]{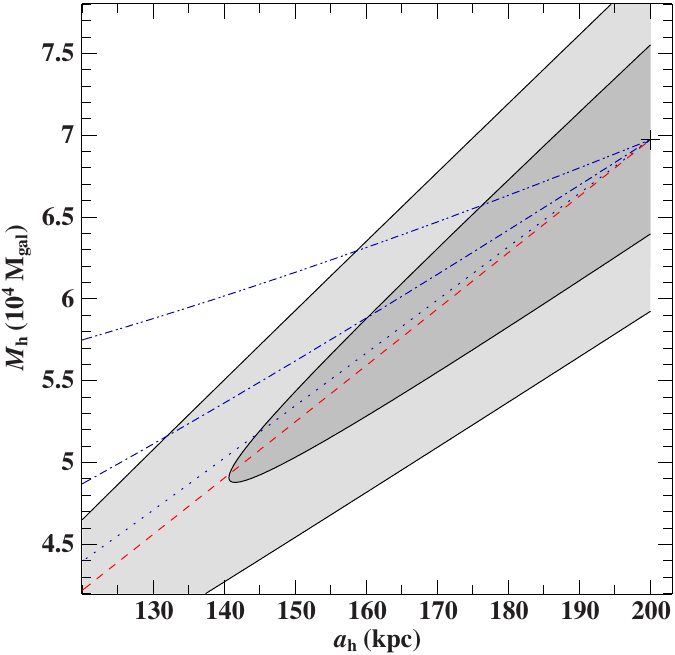}
\caption{Visualization of the correlation between $M_{\mathrm{h}}$ and $a_{\mathrm{h}}$ in Model~II. The meaning of the cross, curves, and shaded regions is the same as in Fig.~\ref{fig:ah_mh_AS} except that the red dashed line defines loci of constant halo mass inside a sphere of radius $20\,\mathrm{kpc}$.}
\label{fig:ah_mh_TF}
\end{figure}
The properties of the best-fitting Model~II are visualized in Figs.~\ref{fig:phi_rho} and \ref{fig:rotation_curve_TF} and summarized in Table~\ref{table:results_TF}. While the rotation curve reveals the same systematic shortcomings as found in Model~I, the match of the remaining constraints is of comparable quality. The unweighted $\chi^2$ per d.o.f. is $1.7$.

Again, there is a tight correlation between the halo parameters $M_{\mathrm{h}}$ and $a_{\mathrm{h}}$ (see Fig.~\ref{fig:ah_mh_TF}). To avoid unphysically large halos, the restriction $a_{\mathrm{h}} \le 200\,\mathrm{kpc}$ has to be imposed based on exactly the same reasoning as the condition $\Lambda \le 200\,\mathrm{kpc}$ in Model~I. Owing to the correlation, this step also sets an upper limit to the parameter $M_{\mathrm{h}}$, hence to the total halo mass $m_{\mathrm{h}}$. Unfortunately, the effects on the latter and thus on the total Galactic mass are more severe in this case since its 90\%-confidence interval, $M_{\mathrm{total}} = 1.7^{+0.2}_{-0.5}\times 10^{12}\mathrm{M}_{\sun}$, covers $a_{\mathrm{h}}$ values only up to $218\,\mathrm{kpc}$. The mass inside $50\,\mathrm{kpc}$, on the other hand, is almost independent of the chosen upper boundary for $a_{\mathrm{h}}$. According to \citet{Sakamoto}, this quantity is much more robust than, e.g., the total mass of the Milky Way. The reason for this is the specific form of the applied potential. It is such that loci of equal masses inside spheres of radii ranging from $\sim$0 to $\sim$50\,kpc lie in a very narrow band in the $M_{\mathrm{h}}$-$a_{\mathrm{h}}$ plane (see Fig.~\ref{fig:ah_mh_TF}). Although the correlation follows contours of equal mass within the central $\sim$20\,kpc, it therefore nearly follows contours of equal mass within the central $50\,\mathrm{kpc}$, resulting in small uncertainties for the latter. The value derived here, $M_{R<50\,\mathrm{kpc}} = 4.6 \pm 0.3 \times 10^{11}\mathrm{M}_{\sun}$, is slightly lower than in \citet{Sakamoto}, $M_{R<50\,\mathrm{kpc}} = 5.4^{+0.1}_{-0.4}\times10^{11}\,\mathrm{M}_{\sun}$. 

In general, the gravitational potential of Model~II is shallower than in Model~I (see Fig.~\ref{fig:phi_rho}) implying systematically lower masses, a lower local escape velocity of about $576\,\mathrm{km}\,\mathrm{s}^{-1}$ in agreement with \citet{smith_etal}, and an unbound orbit for J1539+0239 ($\chi_{*} = 0.80$). Nevertheless, Leo~I is still on a bound orbit.
\section{Model~III}\label{section:NFW}
The third halo potential considered here is based on the universal density profile of dark matter halos suggested by \citet{NFW} from cosmological simulations.
\subsection{Components and characteristics}\label{section:NFW_cc}
Using
\begin{equation}
\rho_{\mathrm{h}}(R) = \frac{M_{\mathrm{h}}}{4 \pi} \frac{1}{\left( a_{\mathrm{h}} + R \right)^2 R}
\label{eq:rho_NFW}
\end{equation}
\citep{NFW} together with Eq.~(\ref{poisson}), one obtains a halo potential of
\begin{equation}
\Phi_{\mathrm{h}}(R) = - \frac{M_{\mathrm{h}}}{R}\ln\left(1+\frac{R}{a_{\mathrm{h}}}\right)\, . \;
\end{equation}
The weighting factor $M_{\mathrm{h}}$ is thereby equivalent to the mass inside a sphere of radius $\sim$5.3 times the scale length $a_{\mathrm{h}}$. The combination of this halo potential with a \citet{Miyamoto_Nagai} disk and bulge component (see Section~\ref{section:bulge_disk}) is hereafter denoted Model~III. Owing to the use of Eq.~(\ref{eq:rho_NFW}) for $R \rightarrow \infty$, the total halo mass is formally logarithmically divergent.
\subsection{Results}\label{section:NFW_Result}
\begin{table*}
\renewcommand{\arraystretch}{1.2}
\caption{\label{table:results_NFW}Properties of the best-fitting Model~III.}
\centering
\begin{tabular}{lc|ccc|cc}
\hline\hline
Parameter & Value & Constraint & \multicolumn{2}{c}{Value} \vline& Derived Quantity & Value\\
\cline{4-5}
          &       &            & Observation & Model       &                  &      \\
\hline
$r_{\sun}\,(\mathrm{kpc})$ & $8.33\pm0.09$ & $r_{\sun}\,(\mathrm{kpc})$ & $8.33 \pm 0.35$ & $8.33$ & $\varv_0\,(\mathrm{km}\,\mathrm{s}^{-1})$ & $239.7$\\
$M_{\mathrm{b}}\,(\mathrm{M}_{\mathrm{gal}})$ & $439\pm28$ & $\mu_{\mathrm{SgrA^*}}\,(\mathrm{mas}\,\mathrm{yr}^{-1})$ & $-6.379 \pm 0.026$ & $-6.380$ & $m_{\mathrm{b}}\,(10^{9}\mathrm{M}_{\sun})$ & $10.2 \pm 0.7$\\
$M_{\mathrm{d}}\,(\mathrm{M}_{\mathrm{gal}})$ & $3096 \pm 197$ & Terminal velocities from \ion{H}{i} & See Fig.~\ref{fig:rotation_curve_NFW} & See Fig.~\ref{fig:rotation_curve_NFW} & $m_{\mathrm{d}}\,(10^{10}\mathrm{M}_{\sun})$ & $7.2 \pm 0.5$\\
$M_{\mathrm{h}}\,(\mathrm{M}_{\mathrm{gal}})$ & $142\,200^{+137\,900}_{-75\,500}$\tablefootmark{a} & Terminal velocities from \element[][]{CO} & See Fig.~\ref{fig:rotation_curve_NFW} & See Fig.~\ref{fig:rotation_curve_NFW} & $m_{\mathrm{h}}\,(10^{12}\mathrm{M}_{\sun})$ & $\infty$\tablefootmark{b}\\
$b_{\mathrm{b}}\,(\mathrm{kpc})$ & $0.236\pm0.021$ & Circular velocities from masers & See Fig.~\ref{fig:rotation_curve_NFW} & See Fig.~\ref{fig:rotation_curve_NFW} & $M_{R<50\,\mathrm{kpc}}\,(10^{12}\mathrm{M}_{\sun})$ & $0.81^{+0.13}_{-0.15}$\\
$a_{\mathrm{d}}\,(\mathrm{kpc})$ & $3.262^{+0.144}_{-0.121}$ & $\rho_{\sun}\,(\mathrm{M}_{\sun}\,\mathrm{pc}^{-3})$, $\Sigma_{1.1}\,(\mathrm{M}_{\sun}\,\mathrm{pc}^{-2})$ & $0.102 \pm 0.010$, $74 \pm 6$ & $0.102$, $75$ & $M_{R<100\,\mathrm{kpc}}\,(10^{12}\mathrm{M}_{\sun})$ & $1.67 \pm 0.46$\\
$b_{\mathrm{d}}\,(\mathrm{kpc})$ & $0.289\pm0.022$ & $\sigma_{\mathrm{BW}}\,(\mathrm{km}\,\mathrm{s}^{-1})$ & $117 \pm 15$ & $123$ & $M_{R<200\,\mathrm{kpc}}\,(10^{12}\mathrm{M}_{\sun})$ & $3.0^{+1.2}_{-1.1}$\\
$a_{\mathrm{h}}\,(\mathrm{kpc})$ & $45.02^{+22.56}_{-16.78}$\tablefootmark{a} & $\chi_{*}$ & $\le 0$ & $0.20$ & $\varv_{\mathrm{esc},\sun}\,(\mathrm{km}\,\mathrm{s}^{-1})$ & $811.5$\\
&  &  &  &  & $A\,(\mathrm{km}\,\mathrm{s}^{-1}\,\mathrm{kpc}^{-1})$ & $14.70$\\
&  &  &  &  & $B\,(\mathrm{km}\,\mathrm{s}^{-1}\,\mathrm{kpc}^{-1})$ & $-14.08$\\ 
\hline
\end{tabular}
\tablefoot{
The quoted uncertainties for the model parameters and masses are 90\%-confidence limits (see notes on Table~\ref{table:results_AS} for details). \tablefoottext{a}{The large uncertainties are due to a strong correlation between $M_{\mathrm{h}}$ and $a_{\mathrm{h}}$, see Fig.~\ref{fig:ah_mh_NFW}.} \tablefoottext{b}{Formal divergence, see Sect.~\ref{section:NFW_cc}.}
}
\end{table*}
\begin{figure*}
\centering
\includegraphics[width=\textwidth]{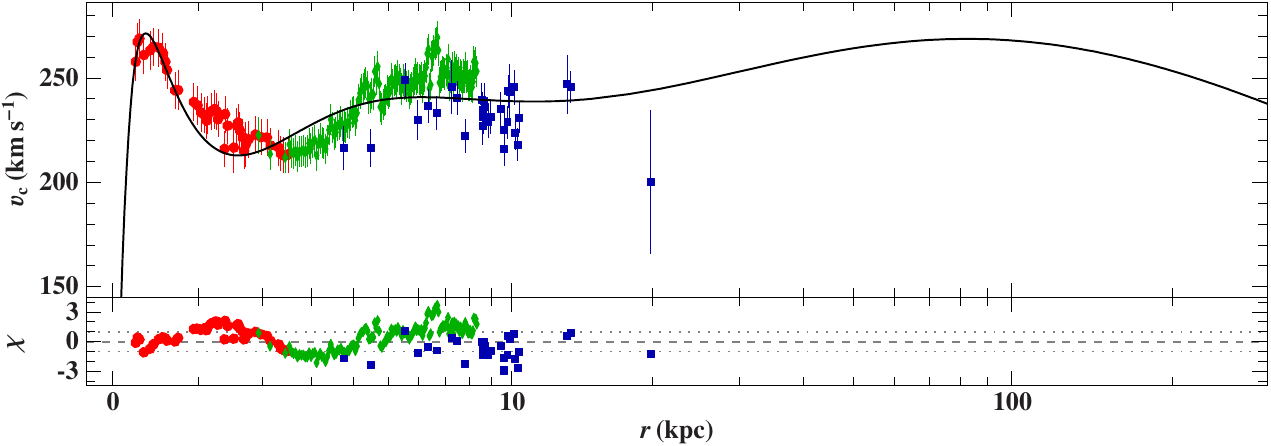}
\caption{Comparison of the best-fitting model rotation curve (solid black line) with terminal velocities from surveys in \ion{H}{i} (red circles) and in \element[][]{CO} (green diamonds), as well as maser observations (blue squares) for Model~III.}
\label{fig:rotation_curve_NFW}
\end{figure*}
\begin{figure}
\centering
\includegraphics[width=0.48\textwidth]{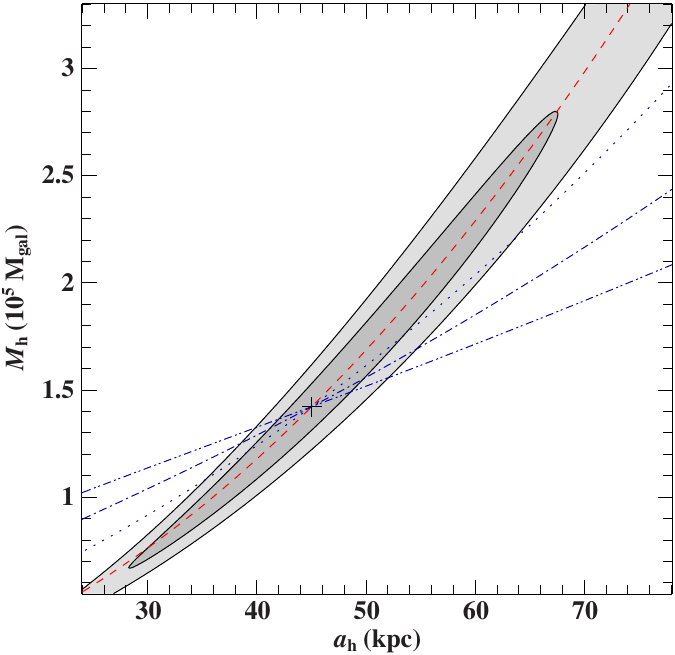}
\caption{Visualization of the correlation between $M_{\mathrm{h}}$ and $a_{\mathrm{h}}$ in Model~III. The meaning of the cross, curves, and shaded regions is the same as in Fig.~\ref{fig:ah_mh_AS} except that the red dashed line defines loci of constant halo mass inside a sphere of radius $18.8\,\mathrm{kpc}$.}
\label{fig:ah_mh_NFW}
\end{figure}
The properties of the best-fitting Model~III are visualized in Figs.~\ref{fig:phi_rho} and \ref{fig:rotation_curve_NFW} and summarized in Table~\ref{table:results_NFW}. The quality of the fit is similar to Models~I and II and yields an unweighted $\chi^2$ per d.o.f. of $1.7$.

In contrast to the two previous models, the rotation curve is rising outside the solar circle, reaching its maximum at about $82\,\mathrm{kpc}$, and falling off beyond this point. This is a consequence of the search for a compromise between the opposing constraints of making J1539+0239 bound, i.e., giving the Milky Way a very high total mass, and, at the same time, limiting the mass in the central $\sim$19\,kpc (see Fig.~\ref{fig:ah_mh_NFW}) to match the remaining observations. Unlike Model~III, Models~I and II were designed to give flat rotation curves, avoiding any analogical behavior. As a result, the masses of Model~III systematically exceed those of the others, e.g., by more than a factor of $1.5$ in terms of $M_{R<50\,\mathrm{kpc}}$. Moreover, because of using Eq.~(\ref{eq:rho_NFW}) for $R \rightarrow \infty$, the total mass $M_{\mathrm{total}}$ is logarithmically infinite. Following \citet{NFW}, an alternative measure of the halo mass is given by $M_{200}$, which is the halo mass inside a sphere of radius $r_{200}$ defined to have a mean interior density of $200$ times the critical value for closure $\rho_{\mathrm{crit}} = 3H^2/8\pi G$. Assuming a Hubble constant $H$ of $73\,\mathrm{km}\,\mathrm{s}^{-1}\mathrm{Mpc}^{-1}$, values of $r_{200}=319^{+61}_{-58}\,\mathrm{kpc}$ and $M_{200} = 4.0^{+1.2}_{-1.8} \times 10^{12}\mathrm{M}_{\sun}$ are derived from Model~III agreeing with $r_{200}=301\,\mathrm{kpc}$ and $M_{200} = 3.4 \times 10^{12}\mathrm{M}_{\sun}$ as used by \citet{Abadi_etal}. The respective local escape velocity of $812\,\mathrm{km}\,\mathrm{s}^{-1}$ significantly exceeds the \citet{smith_etal} value but is comparable to the one in \citet{Abadi_etal}. Although this is high enough to keep Leo~I formally bound, it does not hold J1539+0239 within $200\,\mathrm{kpc}$ ($\chi_{*} = 0.20$).

\citet{kenyon_etal} considered a Galactic potential that only differs from Model~III in the bulge component: the expression $R + b_{\mathrm{b}}$ is used instead of $\sqrt{R^2+{b_{\mathrm{b}}}^2}$. Putting special emphasis on the very central region of the Milky Way, the parameters of their model were tweaked to provide a reasonable match to observations from $R=5\,\mathrm{pc}$ to $R=100\,\mathrm{kpc}$. Compared to Model~III, their resulting model is less massive ($M_{200} = 1.6 \times 10^{12}\mathrm{M}_{\sun}$), which is also reflected by lower values for the local escape velocity ($\varv_{\mathrm{esc},\sun}=635\,\mathrm{km}\,\mathrm{s}^{-1}$), the local mass and surface density ($\rho_{\sun} = 0.046\,\mathrm{M}_{\sun}\,\mathrm{pc}^{-3}$, $\Sigma_{1.1}=46\,\mathrm{M}_{\sun}\,\mathrm{pc}^{-2}$), and the circular motion of the LSR ($\varv_0 = 220\,\mathrm{km}\,\mathrm{s}^{-1}$). 
\section{The hypervelocity star HE~0437-5439 revisited}\label{HE04375439}
\begin{table}
\renewcommand{\arraystretch}{1.2}
\caption{\label{table:input}Kinematic input parameters.}
\centering
\begin{tabular}{lcc}
\hline\hline
Parameter & \multicolumn{2}{c}{Value}\\
\cline{2-3}
& HE~0437-5439 & LMC \\
\hline
$\alpha$~(J2000.0) & $4^{\mathrm{h}}38^{\mathrm{m}}12\fs8$\tablefootmark{a} & $5^{\mathrm{h}}27\fm6$\tablefootmark{d}\\
$\delta$~(J2000.0) & $-54\degr33\arcmin12\arcsec$\tablefootmark{a} & $-69\degr52\arcmin$\tablefootmark{d}\\
distance\,(kpc) & $61 \pm 9$\tablefootmark{b} & $50.1 \pm 2.4$\tablefootmark{e}\\
$\mu_{\alpha}\cos\delta\,(\mathrm{mas} \,\mathrm{yr}^{-1})$ & $0.53 \pm 0.25\mathrm{(stat)} \pm 0.33\mathrm{(sys)}$\tablefootmark{c} & $2.03 \pm 0.08$\tablefootmark{f}\\
$\mu_{\delta}\,(\mathrm{mas} \,\mathrm{yr}^{-1})$ & $0.09 \pm 0.21\mathrm{(stat)} \pm 0.48\mathrm{(sys)}$\tablefootmark{c} & $0.44 \pm 0.05$\tablefootmark{f}\\
$\varv_{\mathrm{rad}}\,(\mathrm{km}\,\mathrm{s}^{-1})$ & $723 \pm 3$\tablefootmark{a} & $262.2 \pm 3.4$\tablefootmark{d}\\
\hline
\end{tabular}
\tablefoot{
Uncertainties are $1\sigma$.
}
\tablebib{
\tablefoottext{a}{\citet{Edelmann_etal}};
\tablefoottext{b}{\citet{Przybilla_etal}};
\tablefoottext{c}{\citet{Brown_etal}};
\tablefoottext{d}{\citet{vanderMarel_etal}};
\tablefoottext{e}{\citet{Freedman_etal}};
\tablefoottext{f}{\citet{kallivayalil_etal}}.
}
\end{table}
\begin{table*}
\renewcommand{\arraystretch}{1.2}
\caption{\label{table:kinematics}Results of the kinematic investigation of HE~0437-5439.}
\centering
\begin{tabular}{lcccccccccccc}
\hline\hline
Run & \multicolumn{3}{c}{Model~I} & & \multicolumn{3}{c}{Model~II} & & \multicolumn{3}{c}{Model~III}\\
\cline{2-4} \cline{6-8} \cline{10-12}
& $d_{\mathrm{p}}\,(\mathrm{kpc})$ & $T_{\mathrm{p}}\,(\mathrm{Myr})$ & $\varv_{\mathrm{p}}\,(\mathrm{km}\,\mathrm{s}^{-1})$ & & $d_{\mathrm{p}}\,(\mathrm{kpc})$ & $T_{\mathrm{p}}\,(\mathrm{Myr})$ & $\varv_{\mathrm{p}}\,(\mathrm{km}\,\mathrm{s}^{-1})$ & & $d_{\mathrm{p}}\,(\mathrm{kpc})$ & $T_{\mathrm{p}}\,(\mathrm{Myr})$ & $\varv_{\mathrm{p}}\,(\mathrm{km}\,\mathrm{s}^{-1})$\\
\hline
\#1a & $13.2 \pm 4.5$ & $-22 \pm 13$ & $640 \pm 70$ & & $13.2 \pm 4.5$ & $-22 \pm 13$ & $640 \pm 70$ & & $13.1 \pm 4.4$ & $-22 \pm 13$ & $640 \pm 70$\\
\#1b & $15.6 \pm 10.4$ & $-98 \pm 15$ & $690 \pm 50$ & & $15.7 \pm 10.3$ & $-98 \pm 16$ & $680 \pm 50$ & & $14.8 \pm 10.1$ & $-95 \pm 14$ & $730 \pm 40$\\
\#2 & $11.0 \pm 4.5$ & $-23 \pm 12$ & $680 \pm 80$ & & $10.9 \pm 4.5$ & $-23 \pm 12$ & $680 \pm 80$ & & $11.0 \pm 4.5$ & $-23 \pm 12$ & $680 \pm 80$\\
\#3a & $7.5 \pm 3.6$ & $-26 \pm 11$ & $680 \pm 80$ & & $7.5 \pm 3.7$ & $-26 \pm 12$ & $680 \pm 80$ & & $7.6 \pm 3.7$ & $-26 \pm 12$ & $680 \pm 80$\\
\#3b & $16.6 \pm 8.6$ & $-95 \pm 12$ & $690 \pm 30$ & & $17.0 \pm 8.9$ & $-95 \pm 12$ & $690 \pm 30$ & & $15.8 \pm 8.5$ & $-93 \pm 12$ & $730 \pm 20$\\
\hline
\end{tabular}
\tablefoot{$d_{\mathrm{p}}$ is the distance, $T_{\mathrm{p}}$ the time, and $\varv_{\mathrm{p}}$ the relative velocity at periastron of HE~0437-5439 with respect to the center of the LMC (runs \#1a, \#2, and \#3a) or to the GC (runs \#1b and \#3b). Numbers are mean value $\pm$ standard deviation $\sigma$. The initial conditions of runs \#1 are those of \citet{Brown_etal}. In runs \#2 and \#3a, \citeauthor{Kroupa_Bastian}'s \citeyearpar{Kroupa_Bastian} LMC proper motions are used. In runs \#3, proper motions of HE~0437-5439 are increased by their systematic errors (CTI correction, see text for details).
}
\end{table*}
\begin{figure*}
\centering
\includegraphics[width=\textwidth]{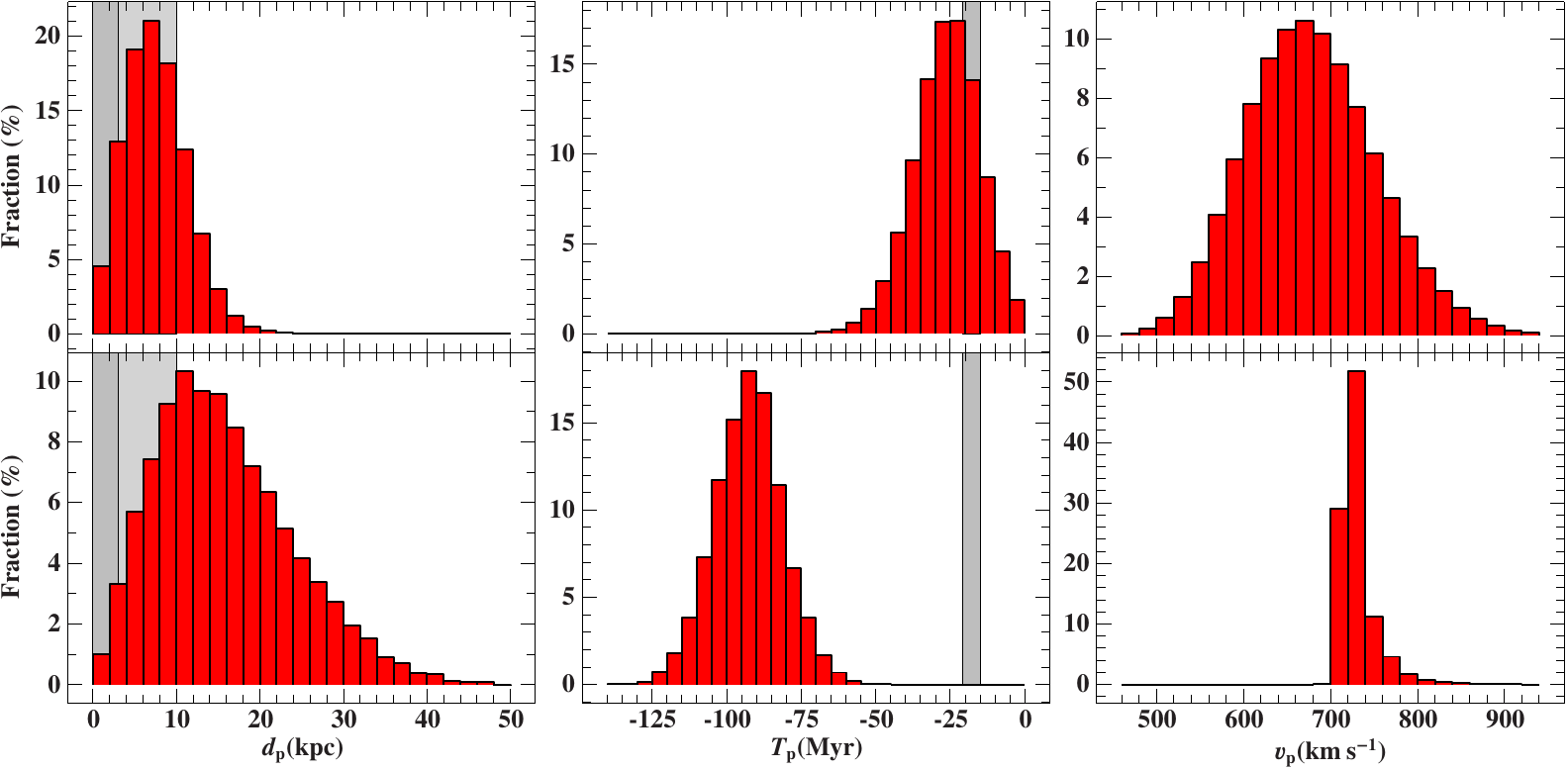}
\caption{Histograms showing the distribution of distances $d_{\mathrm{p}}$, times $T_{\mathrm{p}}$, and relative velocities $\varv_{\mathrm{p}}$ at periastron of HE~0437-5439 with respect to the LMC (\textit{upper panel}, run \#3a) and the GC (\textit{lower panel}, run \#3b) for Model~III. Mean values and standard deviations are given in Table~\ref{table:kinematics}. The gray-shaded areas mark regions with $d_{\mathrm{p}} \le R^{\mathrm{LMC}}_{\mathrm{cen}} = 3\,\mathrm{kpc}$, $d_{\mathrm{p}} \le R^{\mathrm{LMC}}_{\mathrm{out}} = 10\,\mathrm{kpc}$, and $15\,\mathrm{Myr} \le |T_{\mathrm{p}}| \le 21\,\mathrm{Myr}$. The last is the lifetime of HE~0437-5439 assuming a single-star nature \citep{Przybilla_etal}.}
\label{fig:histograms}
\end{figure*}
Given the smoothness of the presented potentials, a simple fourth-order Runge-Kutta method with adaptive stepsize control is sufficient to reliably and efficiently solve the equations of motion (as given in Appendix~\ref{appendix:eom}) numerically. Various trial trajectories, including those of the LSR, were successfully calculated in order to test the self-written ISIS-implementation of the adopted integrator. The hypervelocity star HE~0437-5439 \citep{Edelmann_etal} is analyzed as a first and very interesting application because the spatial origin of this star is still under debate. 

When hypervelocity stars were first discovered in 2005 \citep{brown_etal_hyper}, dynamical ejection from the supermassive black hole at the GC \citep{Hills} was supposed to be their only origin because these stars move so fast that they are unbound to the Galaxy. This scenario was challenged by the discovery of the third hypervelocity star HE~0437-5439 \citep{Edelmann_etal}, a massive B star, because the travel time from the GC to its present position would exceed its lifetime by a factor of 4. \citet{Edelmann_etal} notice that the star is much closer to the Large Magellanic Cloud (LMC) than to the Galaxy and might therefore originate in the LMC. This was corroborated by a sophisticated differential abundance analysis of high-quality, high-resolution spectra by \citet{Przybilla_etal}, who shows that the abundance pattern is inconsistent with that of GC stars but consistent with that of LMC B stars to within error limits. A kinematic investigation for the place of birth of HE~0437-5439 would require precise measurements of the star's, as well as LMC's, proper motion. While several studies have been carried out to measure the proper motion of the LMC, no measurements have been available for HE~0437-5439 until recently. \citet{Brown_etal} used two epochs of images taken with the Advanced Camera for Surveys (ACS) onboard HST to determine the proper motion of HE~0437-5439 and argue that their measurement rules out a place of birth in the LMC at the $3\sigma$~level but is consistent with an origin from the GC. To remedy the time-of-flight versus lifetime problem, HE~0437-5439 needs to be a blue straggler \citep{Edelmann_etal}. This would imply that the progenitor was a triple system from which a binary was ejected by the supermassive black hole. Later on its trajectory, the binary system merged to form the blue straggler \citep{perets}. However, the result of \citet{Brown_etal} is strongly affected by the treatment of LMC motions, as well as the charge-transfer inefficiency (CTI) correction applied to the data. To demonstrate both issues, the kinematical analysis of HE~0437-5439 is revisited here considering three different cases.
\begin{enumerate}
\item To start with, trajectories for $10^4$ Monte-Carlo realizations, which simultaneously account for statistical and systematic uncertainties in current positions and velocities via Gaussian distribution, are computed for HE~0437-5439 and LMC independently using the input parameters listed in Table~\ref{table:input}. The distributions of pericenter distance $d_{\mathrm{p}}$, its corresponding point in time $T_{\mathrm{p}}$, and relative velocity $\varv_{\mathrm{p}}$ derived from the resulting sample of $10^8$ combinations of orbit pairs are insensitive to the choice of the mass model (see run \#1a in Table~\ref{table:kinematics}), and they confirm the results of \citet{Brown_etal} ($d_{\mathrm{p}} = 13\,\mathrm{kpc}$, $T_{\mathrm{p}} = -23\,\mathrm{Myr}$) based on the \citet{kenyon_etal} Galactic potential and identical input values. Since $d_{\mathrm{p}}$ exceeds the radii of LMC's central region $R^{\mathrm{LMC}}_{\mathrm{cen}} = 3\,\mathrm{kpc}$, as well as the outermost regions $R^{\mathrm{LMC}}_{\mathrm{out}} = 10\,\mathrm{kpc}$ \citep{Brown_etal}, an origin in the LMC seems unlikely in this context.

\item Next, the LMC proper motions of Table~\ref{table:input} are replaced by $(\mu_{\alpha}\cos\delta, \mu_{\delta}) = (+1.94 \pm 0.29, -0.14 \pm 0.36)\,\mathrm{mas} \,\mathrm{yr}^{-1}$ \citep{Kroupa_Bastian} to explore their influence on the outcome. Performing the same Monte-Carlo method as before, all three models give smaller pericenter distances (see run \#2 in Table~\ref{table:kinematics}) and thus closer encounters of the two objects.

\item Assuming the CTI of ACS to be rising linearly in time, \citet{Brown_etal} applied 55\% of the epoch-2 CTI correction to their epoch-1 images. According to \citet{massey}, however, there was a dramatic increase in the CTI between the two epochs that lead to an overcorrection in the epoch-1 data by \citeauthor{Brown_etal} implying proper motions of HE~0437-5439 that were larger than stated in \citet{Brown_etal}. The corresponding effects are roughly estimated here by adding the systematic errors to the mean value and omitting them afterwards, i.e., by using $(\mu_{\alpha}\cos\delta, \mu_{\delta}) = (+0.86 \pm 0.25, +0.57 \pm 0.21)\,\mathrm{mas} \,\mathrm{yr}^{-1}$ as input for the Monte-Carlo calculation. The resulting distributions (see run \#3a in Table~\ref{table:kinematics}) are still almost model independent and visualized for Model~III in the upper panel of Fig.~\ref{fig:histograms}. In all three models, about 10\% of all orbit pairs yield pericenter passages within the central region, i.e., $d_{\mathrm{p}} \le R^{\mathrm{LMC}}_{\mathrm{cen}}$, or 76\% in the outermost regions of the LMC ($d_{\mathrm{p}} \le R^{\mathrm{LMC}}_{\mathrm{out}}$). Out of these, 2\% (21\%) have shorter flight times than the star's age of $18 \pm 3\,\mathrm{Myr}$ \citep{Przybilla_etal}. Thus, 0.2\% (16\%) of the trajectories are consistent with an origin in the inner (outer) LMC without invoking additional requirements, such as a blue straggler nature. The decrease in pericenter distances due to CTI effects (Table~\ref{table:kinematics}: run \#2 versus run \#3a) is stronger than when they are due to a change in LMC proper motions (Table~\ref{table:kinematics}: run \#1a versus run \#2). 
\end{enumerate}

For comparison, we now consider the GC as the place of origin. Using the input values of Table~\ref{table:kinematics}, HE~0437-5439 passed $95 \pm 14 \,\mathrm{Myr}$ ago within $d_{\mathrm{p}} = 14.8 \pm 10.1\,\mathrm{kpc}$ of the GC (see Model~III of run \#1b in Table~\ref{table:kinematics}). While the travel time is in good agreement with \citet{Brown_etal} ($T_{\mathrm{p}} = -98\,\mathrm{Myr}$), it is difficult to compare $d_{\mathrm{p}}$ since it is not given in \citet{Brown_etal}. Nevertheless, the upper lefthand panel of Fig.~3 in \citet{Brown_etal} suggests that substantially more than 68\% of all trajectories have Galactic plane-crossing locations below 15\,kpc. Because the Galactic plane-crossing location is an upper limit for $d_{\mathrm{p}}$, this fraction increases further when we consider the pericenter distance. However, this contradicts our result that less than 60\% of all orbits pass the GC within 15\,kpc. As seen from run \#3b in Table~\ref{table:kinematics}, the situation does not change significantly when accounting for the CTI overcorrection. The respective distributions from the $10^4$ trajectories of HE~0437-5439 are shown in the lower panel of Fig.~\ref{fig:histograms}, whereby only $\sim$2\% (27\%) of the trajectories have pericenter passages within 3\,kpc (10\,kpc). All of them exceed the stellar lifetime and therefore require HE~0437-5439 to be a blue straggler.

As a consequence, ruling out an origin in the LMC in favor of the GC is disputable and the question of HE~0437-5439's birthplace remains open.
\section{Conclusions}\label{section:conclusions}
The motions of stars provide important information about the mass distribution in various components of the Galaxy. In particular, they trace the dark matter. \citeauthor{oort}'s \citeyearpar{oort} measurements of stellar motions were the first to hint at the presence of dark matter in the Milky Way. Soon, the \textit{Gaia} mission will provide velocity information of a huge number of stars and satellite galaxies with unprecedented precision. Analytical mass models of the Galaxy are utilized to calculate the orbits of stars. \citet{Allen_Santillan} derived such a model by making use of observational constraints such as the Galactic rotation curve, the distance to the GC, and the local mass density. During the past twenty years, observational data have greatly improved. Therefore, it was time to revisit the Galactic gravitational potential of \citet{Allen_Santillan} and update its parameters by making use of the latest observations. The halo mass is hereby constrained by the most extreme halo star discovered so far \citep{Przybilla_etal}. For comparison, two other widely used halo mass distributions --~the truncated, flat rotation curve model \citep{wilkinson_evans} and a model derived from numerical cosmological simulations \citep{NFW}~-- are fitted as well. 

All three Milky Way mass models are analytical, simple, and equally capable of reproducing their imposed restrictions. Major discrepancies only become apparent at Galactocentric distances greater than $\sim$15\,kpc, which is the region where observational constraints are rare and the halo component dominates. In particular, the depth of the gravitational potential is very sensitive to the form of the dark matter halo seriously affecting for instance predictions of the local escape velocity. The significantly different shapes of the halos allow systematic uncertainties in kinematic investigations to be estimated by comparing the results of orbit computations performed in the three models separately.

Finally, the enigmatic hypervelocity star HE~0437-5439 is re-investigated by inspecting its trajectory in the context of the updated potentials. There are several pros and cons of whether the object originated in the LMC as proposed by the discoverers of the star \citep{Edelmann_etal} or from the GC, which is the suspected place of origin of all hypervelocity stars \citep{brown_etal_hyper}. The problem with the latter is that the travel time to the GC exceeds the stellar lifetime and that the chemical composition of HE~0437-5439 differs from what is known about stars in the GC. A birthplace in the LMC, on the other hand, would be consistent with both. However, based on their own measurement of the star's proper motion, \citet{Brown_etal} claim that the LMC --~in contrast to the GC~-- is ruled out as the place of origin. Therefore, we calculated trajectories for the star by making use of the three mass models under study. Using the same input data as \citet{Brown_etal}, we could confirm their results regarding the LMC, while those for the GC show discrepancies that are independent of the applied mass model. Our trajectories pass the GC within a considerably larger distance, showing that an origin in the GC is much less likely than suggested by \citet{Brown_etal}. Moreover, we investigated the influence of LMC proper motions and inspected systematic errors in the star's proper motion measurements stemming from the CTI overcorrection in the ACS images. The latter have not been considered by \citet{Brown_etal} but turned out to be crucial. Our calculations show that an origin of HE~0437-5439 in the LMC is at least as likely or unlikely as an origin in the GC. Presently available proper motions are therefore inconclusive, and improved measurements are mandatory for settling this issue. 
\begin{acknowledgements}
A.I.~acknowledges support from a research scholarship by the Elite Network of Bavaria. B.W.~and E.T.~acknowledge support for a research internship by the German Academic Exchange Service. L.S.'s research internship was made possible by support from the MIT-Germany Program. This research has made use of NASA's Astrophysics Data System Bibliographic Services. We thank John E.\ Davis for the development of the {\sc slxfig} module used to prepare the figures in this paper, Yoshiaki Sofue for providing the \ion{H}{i} terminal velocity data in digitized form, and Moritz B\"ock for helpful discussions. We are very grateful to Norbert Przybilla and Ulrich Heber for valuable discussions in the course of this work and for their suggestions, comments, and contributions to the manuscript. 
\end{acknowledgements}

\appendix
\section{Equations of motion}\label{appendix:eom}
The motion of test particles in an axisymmetric gravitational potential $\Phi(r,z)$ can be conveniently derived from the Lagrangian~$\mathcal{L}$ of the system:
\begin{equation}
\mathcal{L}\left(r, z, \dot{r}, \dot{\phi}, \dot{z}\right) = \frac{1}{2}\left({\dot{r}}^2 + \left(r \dot{\phi}\right)^2 +{\dot{z}}^2\right) - \Phi(r,z)\, .\;
\end{equation}
Introducing canonical momenta $p_{r} = {\partial \mathcal{L}}/{\partial \dot{r}} = \dot{r}$, $p_{\phi} = {\partial \mathcal{L}}/{\partial \dot{\phi}} = r^2 \dot{\phi}$, and $p_{z} = {\partial \mathcal{L}}/{\partial \dot{z}} = \dot{z}$, the corresponding Lagrange's equations --~written as a system of first-order differential equations~-- read as
\begin{equation}
\dot{r} = p_{r}\, , \;
\end{equation}
\begin{equation}
\dot{\phi} = \frac{p_{\phi}}{r^2}\, , \;
\end{equation}
\begin{equation}
\dot{z} = p_{z}\, , \;
\end{equation}
\begin{eqnarray}
\dot{p}_{r} & = & \frac{{p_{\phi}}^2}{r^3} -\frac{G M_{\mathrm{b}} r}{\left(r^2+z^2+{b_{\mathrm{b}}}^2\right)^{3/2}} - \frac{G M_{\mathrm{d}} r}{\left(r^2 + \left({a_{\mathrm{d}}} + \sqrt{z^2 +{b_{\mathrm{d}}}^2}\right)^2\right)^{3/2}}\nonumber \\& & - \frac{\partial \Phi_{\mathrm{h}}(R)}{\partial R}\frac{r}{\sqrt{r^2+z^2}}\, , \;
\end{eqnarray}
%\begin{subnumcases}{}
%- \frac{G M_{\mathrm{h}} r}{(r^2+z^2) a_{\mathrm{h}}} \frac{\left(\frac{\sqrt{r^2+z^2}}{a_{\mathrm{h}}}\right)^{\gamma-1}}{\left(1+\left(\frac{\sqrt{r^2+z^2}}{a_{\mathrm{h}}}\right)^{\gamma-1}\right)} & \hspace{-2ex} if $\sqrt{r^2+z^2} < \Lambda$,\\
%- \frac{G M_{\mathrm{h}} r}{\left(r^2+z^2\right)^{3/2}} \frac{\left(\frac{\Lambda}{a_{\mathrm{h}}}\right)^{\gamma}}{1+\left(\frac{\Lambda}{a_{\mathrm{h}}}\right)^{\gamma-1}} & \hspace{-2ex} else,
%\end{subnumcases}
\begin{equation}
\dot{p}_{\phi} = 0\, , \;
\end{equation}
\begin{eqnarray}
\dot{p}_{z} & = & -\frac{G M_{\mathrm{b}} z}{\left(r^2+z^2+{b_{\mathrm{b}}}^2\right)^{3/2}} - \frac{G M_{\mathrm{d}} z \left(\frac{a_{\mathrm{d}}}{\sqrt{z^2 +{b_{\mathrm{d}}}^2}}+1\right)}{\left(r^2 + \left({a_{\mathrm{d}}} + \sqrt{z^2 +{b_{\mathrm{d}}}^2}\right)^2\right)^{3/2}}\nonumber \\& & - \frac{\partial \Phi_{\mathrm{h}}(R)}{\partial R}\frac{z}{\sqrt{r^2+z^2}}\, , \;
\end{eqnarray}
%\begin{subnumcases}{}
%- \frac{G M_{\mathrm{h}} z}{(r^2+z^2) a_{\mathrm{h}}} \frac{\left(\frac{\sqrt{r^2+z^2}}{a_{\mathrm{h}}}\right)^{\gamma-1}}{\left(1+\left(\frac{\sqrt{r^2+z^2}}{a_{\mathrm{h}}}\right)^{\gamma-1}\right)} & \hspace{-2ex} if $\sqrt{r^2+z^2} < \Lambda$,\\
%- \frac{G M_{\mathrm{h}} z}{\left(r^2+z^2\right)^{3/2}} \frac{\left(\frac{\Lambda}{a_{\mathrm{h}}}\right)^{\gamma}}{1+\left(\frac{\Lambda}{a_{\mathrm{h}}}\right)^{\gamma-1}} & \hspace{-2ex} else.
%\end{subnumcases}
whereby
\[
\frac{\partial \Phi_{\mathrm{h}}(R)}{\partial R}=
\]
\begin{subnumcases}{}
\frac{G M_{\mathrm{h}}}{a_{\mathrm{h}}R} \frac{\left(\frac{R}{a_{\mathrm{h}}}\right)^{\gamma-1}}{1+\left(\frac{R}{a_{\mathrm{h}}}\right)^{\gamma-1}} & \hspace{-2ex} if $R < \Lambda$ \& Model~I,\\
\frac{G M_{\mathrm{h}}}{R^2} \frac{\left(\frac{\Lambda}{a_{\mathrm{h}}}\right)^{\gamma}}{1+\left(\frac{\Lambda}{a_{\mathrm{h}}}\right)^{\gamma-1}} & \hspace{-2ex} if $R \ge \Lambda$ \& Model~I,\\
  \frac{G M_{\mathrm{h}}}{\sqrt{R^2+{a_{\mathrm{h}}}^2} R} & \hspace{-2ex} if Model~II,\\
  \frac{G M_{\mathrm{h}}}{R^2} \ln{\left(1+\frac{R}{a_{\mathrm{h}}}\right)} - \frac{G M_{\mathrm{h}}}{(a_{\mathrm{h}}+R)R } & \hspace{-2ex} if Model~III.
\end{subnumcases}
Expressing lengths in $\mathrm{kpc}$, masses in galactic mass units $\mathrm{M}_{\mathrm{gal}}$, and time in $\mathrm{Myr}$, the gravitational constant $G$ has to be replaced by the factor $\sim\!\!\!1.04598 \times 10^{-4}$ thereby converting units from $100\,\mathrm{km}^2\,\mathrm{s}^{-2}$ to $\mathrm{kpc}^2\,\mathrm{Myr}^{-2}$.

\bibliographystyle{aa}
%\bibliography{bib_paper}

\begin{thebibliography}{64}
\expandafter\ifx\csname natexlab\endcsname\relax\def\natexlab#1{#1}\fi

\bibitem[{{Abadi} {et~al.}(2009){Abadi}, {Navarro}, \&
  {Steinmetz}}]{Abadi_etal}
{Abadi}, M.~G., {Navarro}, J.~F., \& {Steinmetz}, M. 2009, \apjl, 691, L63

\bibitem[{{Allen} \& {Martos}(1986)}]{Allen_Martos}
{Allen}, C. \& {Martos}, M.~A. 1986, \rmxaa, 13, 137

\bibitem[{{Allen} {et~al.}(2008){Allen}, {Moreno}, \&
  {Pichardo}}]{allen_moreno_pichardo}
{Allen}, C., {Moreno}, E., \& {Pichardo}, B. 2008, \apj, 674, 237

\bibitem[{{Allen} \& {Santill\'{a}n}(1991)}]{Allen_Santillan}
{Allen}, C. \& {Santill\'{a}n}, A. 1991, \rmxaa, 22, 255

\bibitem[{{Altmann} \& {de Boer}(2000)}]{altmann_deboer}
{Altmann}, M. \& {de Boer}, K.~S. 2000, \aap, 353, 135

\bibitem[{{Altmann} {et~al.}(2004){Altmann}, {Edelmann}, \& {de
  Boer}}]{altmann_edelmann_deboer}
{Altmann}, M., {Edelmann}, H., \& {de Boer}, K.~S. 2004, \aap, 414, 181

\bibitem[{{Asaki} {et~al.}(2010){Asaki}, {Deguchi}, {Imai}, {Hachisuka},
  {Miyoshi}, \& {Honma}}]{masers7}
{Asaki}, Y., {Deguchi}, S., {Imai}, H., {et~al.} 2010, \apj, 721, 267

\bibitem[{{Boylan-Kolchin} {et~al.}(2012){Boylan-Kolchin}, {Bullock}, {Sohn},
  {Besla}, \& {van der Marel}}]{leoimass}
{Boylan-Kolchin}, M., {Bullock}, J.~S., {Sohn}, S.~T., {Besla}, G., \& {van der
  Marel}, R.~P. 2012, \apj, 768, 140

\bibitem[{{Brown} {et~al.}(2010){Brown}, {Anderson}, {Gnedin}, {Bond},
  {Geller}, {Kenyon}, \& {Livio}}]{Brown_etal}
{Brown}, W.~R., {Anderson}, J., {Gnedin}, O.~Y., {et~al.} 2010, \apjl, 719, L23

\bibitem[{{Brown} {et~al.}(2005){Brown}, {Geller}, {Kenyon}, \&
  {Kurtz}}]{brown_etal_hyper}
{Brown}, W.~R., {Geller}, M.~J., {Kenyon}, S.~J., \& {Kurtz}, M.~J. 2005,
  \apjl, 622, L33

\bibitem[{{Burton} \& {Gordon}(1978)}]{Burton_Gordon}
{Burton}, W.~B. \& {Gordon}, M.~A. 1978, \aap, 63, 7

\bibitem[{{Clemens}(1985)}]{Clemens}
{Clemens}, D.~P. 1985, \apj, 295, 422

\bibitem[{{Dehnen} \& {Binney}(1998)}]{Dehnen_Binney}
{Dehnen}, W. \& {Binney}, J. 1998, \mnras, 294, 429

\bibitem[{{Edelmann} {et~al.}(2005){Edelmann}, {Napiwotzki}, {Heber},
  {Christlieb}, \& {Reimers}}]{Edelmann_etal}
{Edelmann}, H., {Napiwotzki}, R., {Heber}, U., {Christlieb}, N., \& {Reimers},
  D. 2005, \apjl, 634, L181

\bibitem[{{Freedman} {et~al.}(2001){Freedman}, {Madore}, {Gibson}, {Ferrarese},
  {Kelson}, {Sakai}, {Mould}, {Kennicutt}, {Ford}, {Graham}, {Huchra},
  {Hughes}, {Illingworth}, {Macri}, \& {Stetson}}]{Freedman_etal}
{Freedman}, W.~L., {Madore}, B.~F., {Gibson}, B.~K., {et~al.} 2001, \apj, 553,
  47

\bibitem[{{Gillessen} {et~al.}(2009){Gillessen}, {Eisenhauer}, {Trippe},
  {Alexander}, {Genzel}, {Martins}, \& {Ott}}]{Gillessen_etal}
{Gillessen}, S., {Eisenhauer}, F., {Trippe}, S., {et~al.} 2009, \apj, 692, 1075

\bibitem[{{Gnedin} {et~al.}(2010){Gnedin}, {Brown}, {Geller}, \&
  {Kenyon}}]{Gnedin_etal}
{Gnedin}, O.~Y., {Brown}, W.~R., {Geller}, M.~J., \& {Kenyon}, S.~J. 2010,
  \apjl, 720, L108

\bibitem[{{Hills}(1988)}]{Hills}
{Hills}, J.~G. 1988, \nat, 331, 687

\bibitem[{{Hirota} {et~al.}(2008{\natexlab{a}}){Hirota}, {Ando}, {Bushimata},
  {Choi}, {Honma}, {Imai}, {Iwadate}, {Jike}, {Kameno}, {Kameya}, {Kamohara},
  {Kan-Ya}, {Kawaguchi}, {Kijima}, {Kim}, {Kobayashi}, {Kuji}, {Kurayama},
  {Manabe}, {Matsui}, {Matsumoto}, {Miyaji}, {Miyazaki}, {Nagayama},
  {Nakagawa}, {Namikawa}, {Nyu}, {Oh}, {Omodaka}, {Oyama}, {Sakai}, {Sasao},
  {Sato}, {Sato}, {Shibata}, {Tamura}, {Ueda}, \& {Yamashita}}]{masers2}
{Hirota}, T., {Ando}, K., {Bushimata}, T., {et~al.} 2008{\natexlab{a}}, \pasj,
  60, 961

\bibitem[{{Hirota} {et~al.}(2008{\natexlab{b}}){Hirota}, {Bushimata}, {Choi},
  {Honma}, {Imai}, {Iwadate}, {Jike}, {Kameya}, {Kamohara}, {Kan-Ya},
  {Kawaguchi}, {Kijima}, {Kobayashi}, {Kuji}, {Kurayama}, {Manabe}, {Miyaji},
  {Nagayama}, {Nakagawa}, {Oh}, {Omodaka}, {Oyama}, {Sakai}, {Sasao}, {Sato},
  {Shibata}, {Tamura}, \& {Yamashita}}]{masers3}
{Hirota}, T., {Bushimata}, T., {Choi}, Y.~K., {et~al.} 2008{\natexlab{b}},
  \pasj, 60, 37

\bibitem[{{Hirsch} {et~al.}(2005){Hirsch}, {Heber}, {O'Toole}, \&
  {Bresolin}}]{hirsch_etal}
{Hirsch}, H.~A., {Heber}, U., {O'Toole}, S.~J., \& {Bresolin}, F. 2005, \aap,
  444, L61

\bibitem[{{Holmberg} \& {Flynn}(2000)}]{Holmberg_Flynn_density}
{Holmberg}, J. \& {Flynn}, C. 2000, \mnras, 313, 209

\bibitem[{{Holmberg} \& {Flynn}(2004)}]{Holmberg_Flynn_surface}
{Holmberg}, J. \& {Flynn}, C. 2004, \mnras, 352, 440

\bibitem[{{Honma} {et~al.}(2011){Honma}, {Hirota}, {Kan-Ya}, {Kawaguchi},
  {Kobayashi}, {Kurayama}, \& {Sato}}]{masers9}
{Honma}, M., {Hirota}, T., {Kan-Ya}, Y., {et~al.} 2011, \pasj, 63, 17

\bibitem[{{Houck} \& {Denicola}(2000)}]{isis}
{Houck}, J.~C. \& {Denicola}, L.~A. 2000, in Astronomical Data Analysis Software and
  Systems IX, eds. {N.~Manset, C.~Veillet, \& D.~Crabtree}, ASP Conf.\ Ser., 216, 591
  
\bibitem[{{Irrgang} {et~al.}(2010){Irrgang}, {Przybilla}, {Heber}, {Nieva}, \&
  {Schuh}}]{irrgang_etal}
{Irrgang}, A., {Przybilla}, N., {Heber}, U., {Nieva}, M.~F., \& {Schuh}, S.
  2010, \apj, 711, 138

\bibitem[{{Kaempf} {et~al.}(2005){Kaempf}, {de Boer}, \&
  {Altmann}}]{kaempf_deboer_altmann}
{Kaempf}, T.~A., {de Boer}, K.~S., \& {Altmann}, M. 2005, \aap, 432, 879

\bibitem[{{Kallivayalil} {et~al.}(2006){Kallivayalil}, {van der Marel},
  {Alcock}, {Axelrod}, {Cook}, {Drake}, \& {Geha}}]{kallivayalil_etal}
{Kallivayalil}, N., {van der Marel}, R.~P., {Alcock}, C., {et~al.} 2006, \apj,
  638, 772

\bibitem[{{Kenyon} {et~al.}(2008){Kenyon}, {Bromley}, {Geller}, \&
  {Brown}}]{kenyon_etal}
{Kenyon}, S.~J., {Bromley}, B.~C., {Geller}, M.~J., \& {Brown}, W.~R. 2008,
  \apj, 680, 312

\bibitem[{{Kroupa} \& {Bastian}(1997)}]{Kroupa_Bastian}
{Kroupa}, P. \& {Bastian}, U. 1997, \na, 2, 77

\bibitem[{{Lane} {et~al.}(2012){Lane}, {K{\"u}pper}, \& {Heggie}}]{lane_etal}
{Lane}, R.~R., {K{\"u}pper}, A.~H.~W., \& {Heggie}, D.~C. 2012, \mnras, 423,
  2845

\bibitem[{{L{\'e}pine} {et~al.}(2011){L{\'e}pine}, {Koch}, {Rich}, \&
  {Kuijken}}]{lepine_etal}
{L{\'e}pine}, S., {Koch}, A., {Rich}, R.~M., \& {Kuijken}, K. 2011, \apj, 741,
  100

\bibitem[{{Massey}(2010)}]{massey}
{Massey}, R. 2010, \mnras, 409, L109

\bibitem[{{McMillan}(2011)}]{McMillan}
{McMillan}, P.~J. 2011, \mnras, 414, 2446

\bibitem[{{McMillan} \& {Binney}(2010)}]{McMillan_Binney}
{McMillan}, P.~J. \& {Binney}, J.~J. 2010, \mnras, 402, 934

\bibitem[{{Miyamoto} \& {Nagai}(1975)}]{Miyamoto_Nagai}
{Miyamoto}, M. \& {Nagai}, R. 1975, \pasj, 27, 533

\bibitem[{{Navarro} {et~al.}(1997){Navarro}, {Frenk}, \& {White}}]{NFW}
{Navarro}, J.~F., {Frenk}, C.~S., \& {White}, S.~D.~M. 1997, \apj, 490, 493

\bibitem[{{Niinuma} {et~al.}(2011){Niinuma}, {Nagayama}, {Hirota}, {Honma},
  {Motogi}, {Nakagawa}, {Kurayama}, {Kan-Ya}, {Kawaguchi}, {Kobayashi}, \&
  {Ueno}}]{masers8}
{Niinuma}, K., {Nagayama}, T., {Hirota}, T., {et~al.} 2011, \pasj, 63, 9

\bibitem[{{Odenkirchen} {et~al.}(1997){Odenkirchen}, {Brosche}, {Geffert}, \&
  {Tucholke}}]{odenkirchen_brosche}
{Odenkirchen}, M., {Brosche}, P., {Geffert}, M., \& {Tucholke}, H.-J. 1997,
  \na, 2, 477

\bibitem[{{Oort}(1932)}]{oort}
{Oort}, J.~H. 1932, \bain, 6, 249

\bibitem[{{Pauli} {et~al.}(2003){Pauli}, {Napiwotzki}, {Altmann}, {Heber},
  {Odenkirchen}, \& {Kerber}}]{pauli1}
{Pauli}, E.-M., {Napiwotzki}, R., {Altmann}, M., {et~al.} 2003, \aap, 400, 877

\bibitem[{{Pauli} {et~al.}(2006){Pauli}, {Napiwotzki}, {Heber}, {Altmann}, \&
  {Odenkirchen}}]{pauli2}
{Pauli}, E.-M., {Napiwotzki}, R., {Heber}, U., {Altmann}, M., \& {Odenkirchen},
  M. 2006, \aap, 447, 173

\bibitem[{{Pereira} {et~al.}(2012){Pereira}, {Jilinski}, {Drake}, {de Castro},
  {Ortega}, {Chavero}, \& {Roig}}]{pereira_etal}
{Pereira}, C.~B., {Jilinski}, E., {Drake}, N.~A., {et~al.} 2012, \aap, 543, A58

\bibitem[{{Perets}(2009)}]{perets}
{Perets}, H.~B. 2009, \apj, 698, 1330

\bibitem[{{Pichardo} {et~al.}(2004){Pichardo}, {Martos}, \&
  {Moreno}}]{pichardo_etal_bar}
{Pichardo}, B., {Martos}, M., \& {Moreno}, E. 2004, \apj, 609, 144

\bibitem[{{Pichardo} {et~al.}(2003){Pichardo}, {Martos}, {Moreno}, \&
  {Espresate}}]{pichardo_etal}
{Pichardo}, B., {Martos}, M., {Moreno}, E., \& {Espresate}, J. 2003, \apj, 582,
  230

\bibitem[{{Przybilla} {et~al.}(2008){Przybilla}, {Nieva}, {Heber}, {Firnstein},
  {Butler}, {Napiwotzki}, \& {Edelmann}}]{Przybilla_etal}
{Przybilla}, N., {Nieva}, M.~F., {Heber}, U., {et~al.} 2008, \aap, 480, L37

\bibitem[{{Przybilla} {et~al.}(2010){Przybilla}, {Tillich}, {Heber}, \&
  {Scholz}}]{J15390239}
{Przybilla}, N., {Tillich}, A., {Heber}, U., \& {Scholz}, R.-D. 2010, \apj,
  718, 37

\bibitem[{{Reid} \& {Brunthaler}(2004)}]{Reid_Brunthaler}
{Reid}, M.~J. \& {Brunthaler}, A. 2004, \apj, 616, 872

\bibitem[{{Reid} {et~al.}(2009){Reid}, {Menten}, {Zheng}, {Brunthaler},
  {Moscadelli}, {Xu}, {Zhang}, {Sato}, {Honma}, {Hirota}, {Hachisuka}, {Choi},
  {Moellenbrock}, \& {Bartkiewicz}}]{masers4}
{Reid}, M.~J., {Menten}, K.~M., {Zheng}, X.~W., {et~al.} 2009, \apj, 700, 137

\bibitem[{{Rygl} {et~al.}(2010){Rygl}, {Brunthaler}, {Reid}, {Menten}, {van
  Langevelde}, \& {Xu}}]{masers5}
{Rygl}, K.~L.~J., {Brunthaler}, A., {Reid}, M.~J., {et~al.} 2010, \aap, 511, A2

\bibitem[{{Sakamoto} {et~al.}(2003){Sakamoto}, {Chiba}, \& {Beers}}]{Sakamoto}
{Sakamoto}, T., {Chiba}, M., \& {Beers}, T.~C. 2003, \aap, 397, 899

\bibitem[{{Sandstrom} {et~al.}(2007){Sandstrom}, {Peek}, {Bower}, {Bolatto}, \&
  {Plambeck}}]{masers1}
{Sandstrom}, K.~M., {Peek}, J.~E.~G., {Bower}, G.~C., {Bolatto}, A.~D., \&
  {Plambeck}, R.~L. 2007, \apj, 667, 1161

\bibitem[{{Sato} {et~al.}(2010){Sato}, {Hirota}, {Reid}, {Honma}, {Kobayashi},
  {Iwadate}, {Miyaji}, \& {Shibata}}]{masers6}
{Sato}, M., {Hirota}, T., {Reid}, M.~J., {et~al.} 2010, \pasj, 62, 287

\bibitem[{{Sch{\"o}nrich} {et~al.}(2010){Sch{\"o}nrich}, {Binney}, \&
  {Dehnen}}]{Schonrich_etal}
{Sch{\"o}nrich}, R., {Binney}, J., \& {Dehnen}, W. 2010, \mnras, 403, 1829

\bibitem[{{Schuster} {et~al.}(2012){Schuster}, {Moreno}, {Nissen}, \&
  {Pichardo}}]{schuster_et_al}
{Schuster}, W.~J., {Moreno}, E., {Nissen}, P.~E., \& {Pichardo}, B. 2012, \aap,
  538, A21

\bibitem[{{Silva} \& {Napiwotzki}(2011)}]{silva_napiwotzki}
{Silva}, M.~D.~V. \& {Napiwotzki}, R. 2011, \mnras, 411, 2596

\bibitem[{{Smith} {et~al.}(2007){Smith}, {Ruchti}, {Helmi}, {Wyse},
  {Fulbright}, {Freeman}, {Navarro}, {Seabroke}, {Steinmetz}, {Williams},
  {Bienaym{\'e}}, {Binney}, {Bland-Hawthorn}, {Dehnen}, {Gibson}, {Gilmore},
  {Grebel}, {Munari}, {Parker}, {Scholz}, {Siebert}, {Watson}, \&
  {Zwitter}}]{smith_etal}
{Smith}, M.~C., {Ruchti}, G.~R., {Helmi}, A., {et~al.} 2007, \mnras, 379, 755

\bibitem[{{Sohn} {et~al.}(2012){Sohn}, {Besla}, {van der Marel},
  {Boylan-Kolchin}, {Majewski}, \& {Bullock}}]{leoi}
{Sohn}, S.~T., {Besla}, G., {van der Marel}, R.~P., {et~al.} 2012, \apj, 768, 139

\bibitem[{{Tillich} {et~al.}(2011){Tillich}, {Heber}, {Geier}, {Hirsch},
  {Maxted}, {G{\"a}nsicke}, {Marsh}, {Napiwotzki}, {{\O}stensen}, \&
  {Scholz}}]{tillich_etal}
{Tillich}, A., {Heber}, U., {Geier}, S., {et~al.} 2011, \aap, 527, A137

\bibitem[{{van der Marel} {et~al.}(2002){van der Marel}, {Alves}, {Hardy}, \&
  {Suntzeff}}]{vanderMarel_etal}
{van der Marel}, R.~P., {Alves}, D.~R., {Hardy}, E., \& {Suntzeff}, N.~B. 2002,
  \aj, 124, 2639

\bibitem[{{Wilkinson} \& {Evans}(1999)}]{wilkinson_evans}
{Wilkinson}, M.~I. \& {Evans}, N.~W. 1999, \mnras, 310, 645

\bibitem[{{Wu} {et~al.}(2011){Wu}, {Ma}, {Zhou}, \& {Du}}]{wu_etal}
{Wu}, Z.-Y., {Ma}, J., {Zhou}, X., \& {Du}, C.-H. 2011, \aj, 141, 104

\bibitem[{{Xue} {et~al.}(2008){Xue}, {Rix}, {Zhao}, {Re Fiorentin}, {Naab},
  {Steinmetz}, {van den Bosch}, {Beers}, {Lee}, {Bell}, {Rockosi}, {Yanny},
  {Newberg}, {Wilhelm}, {Kang}, {Smith}, \& {Schneider}}]{xue_etal}
{Xue}, X.~X., {Rix}, H.~W., {Zhao}, G., {et~al.} 2008, \apj, 684, 1143

\end{thebibliography}

\end{document}